\documentclass[authoryear,preprint,12pt]{elsarticle}

\usepackage{amssymb}
\usepackage{amsmath}

\usepackage{natbib}
\usepackage{longtable}
\usepackage{listings}
\usepackage{xcolor}
\usepackage{booktabs}
\usepackage{threeparttable}
\usepackage{multirow}
\usepackage{url}
\usepackage{hyperref}
\usepackage{float}
\usepackage{placeins}

\definecolor{codebg}{rgb}{0.97,0.97,0.97}
\definecolor{purple}{RGB}{111,0,255}
\definecolor{gray}{RGB}{102,102,102}
\definecolor{lightblue}{RGB}{0,102,204}
\definecolor{darkgreen}{rgb}{0.0, 0.5, 0.0}

\newcommand{\rqtakeaway}[1]{%
  \begin{center}
    {\setlength{\fboxsep}{8pt}%
    \fbox{\parbox{0.93\linewidth}{\textbf{Takeaway.} #1}}}
  \end{center}
}

\lstdefinelanguage{Python}{
    morekeywords={
        def,return,if,elif,else,for,while,break,continue,pass,
        in,is,not,and,or,import,from,as,class,with,lambda,try,except,finally,
        assert,raise,yield,del,global,nonlocal,
        True,False,None,test_,pytest
    },
    sensitive=true,
    morecomment=[l]\#,
    morestring=[b]',
    morestring=[b]",
}

\journal{Journal of Systems and Software}

\begin{document}

\begin{frontmatter}



\title{Exploring Test Smells Across Programming Languages: A Systematic Mapping Study}



\author{Publio Silva\textsuperscript{a}, Carla Bezerra\textsuperscript{b}, Ivan Machado\textsuperscript{b}, Márcio Ribeiro\textsuperscript{c}}

\cortext[cor1]{Corresponding author: Publio Silva, email: publioufc@alu.ufc.br}

\affiliation[a]{
    organization={Federal University of Ceará},
    addressline={Av. José de Freitas Queiroz, 5003},
    city={Quixadá},
    postcode={63900-000},
    state={Ceará},
    country={Brazil}
}

\affiliation[b]{
    organization={Federal University of Bahia},
    addressline={Av. Milton Santos},
    city={Salvador},
    postcode={40170-110},
    state={Bahia},
    country={Brazil}
}

\affiliation[c]{
    organization={Federal University of Alagoas},
    addressline={Av. Lourival Melo Mota},
    city={Maceió},
    postcode={57072-970},
    state={Alagoas},
    country={Brazil}
}

\begin{abstract}
Tests are essential for ensuring code quality in software development. However, poor implementation practices can compromise the maintainability and evolution of test code, leading to additional costs and effort. These practices, known as test smells, have been the focus of numerous studies that expanded the catalog of known test smells. Despite this progress, little attention has been given to examining test smells from the perspective of programming languages, particularly their relationship with specific languages or test frameworks. As a result, researchers and practitioners have proposed various approaches and tools to detect and refactor the same test smells across different languages. In this work, we address this gap by analyzing test smells in the literature from a programming language perspective. We conducted a systematic mapping study, gathering data from papers published up to August 2025, examining 117 papers. We identified the most prevalent languages and frameworks associated with test smells, the most common test smells in each language, refactorings for addressing test smells, datasets of test smells, the criticality of specific test smells, and a collection of tools to detect or refactor test smells for various languages and frameworks. Our results show that while many test smells have been addressed, research remains limited to a few languages and frameworks. We also found datasets and detection/refactoring tools for some languages and analyzed the criticality of frequently mentioned test smells. This study highlights the need to expand research to more languages and develop more universal solutions.
\end{abstract}

\begin{graphicalabstract}
\end{graphicalabstract}

\begin{highlights}
    \item Java is the most studied language in the context of test smells.
    \item \textit{Assertion Roulette} and \textit{Eager Test} are the most frequently cited test smells.
    \item We identified 94 refactorings for 50 different test smells.
    \item Most critical smells: \textit{Sleepy Test}, \textit{Ignored Tests}, \textit{Resource Optimism}, \textit{Mystery Guest}.
    \item We found 31 test smell related tools, mostly designed for Java.
\end{highlights}

\begin{keyword}


test smell \sep programming language \sep test framework
\end{keyword}

\end{frontmatter}



\section{Introduction}\label{sec:introduction}

Testing stands as one of the fundamental practices for ensuring code quality in software development~\citep{Tran2021}. While manual testing requires human intervention, automated testing offers a more efficient alternative through test scripts that provide repeatability with significantly reduced effort~\citep{Garousi2018}. For organizations embracing continuous delivery, automated testing becomes particularly crucial, enabling them to accelerate the journey from feature conception to customer delivery while maintaining high quality standards~\citep{Tran2021}.

However, just as production code must evolve to remain relevant, test code requires similar attention to maintain its effectiveness over time~\citep{Tran2021}. The consequences of poorly implemented test code extend beyond mere inefficiency, they translate into significant extra costs and unnecessary effort~\citep{Garousi2018}. For example, consider the \textit{Assertion Roulette} test smell, which consists of multiple assertions within a single test without explanatory messages. This practice can significantly increase debugging effort, as a failure in one assertion forces the developer to inspect each assertion individually to identify the source of the problem. Another example is the \textit{Conditional Test Logic} test smell, which increases test complexity. Depending on the number of conditions introduced, it can make tests harder to understand, maintain, and update.

This concern becomes particularly pressing when we consider that developers dedicate approximately one-quarter of their time to engineering test code~\citep{Beller2015}. Among the various factors that can compromise the maintainability and evolution of test code, test smells emerge as a critical concern. Test smells represent poor design choices made by developers during test implementation~\citep{Aljedaani2021}, and these choices can ultimately result in test code that becomes difficult to maintain and evolve~\citep{Panichella2022}.

The foundation of test smell research was established by~\citet{Deursen2001}, who introduced an initial catalog of 11 test smells along with their corresponding refactoring strategies. This seminal work has since inspired a wealth of research, with numerous studies expanding the field by proposing new test smells, developing refactoring techniques, and creating detection and refactoring tools. The research landscape has been further enriched by comprehensive literature reviews that have compiled extensive catalogs of test smells~\citep{Garousi2018, Garousi2019}, systematically identified and analyzed test smell detection tools~\citep{Aljedaani2021}, and explored test smells within specific domains~\citep{Rwemalika2022}. Despite this substantial body of work, a critical gap remains: none of these studies specifically examines the prevalence of test smells across different programming languages and frameworks. Consequently, fundamental questions remain unanswered, such as which test smells are language-specific and which are more prevalent in particular programming environments.

Overlooking programming language perspectives in test smell research has created a significant practical challenge. Many studies propose state-of-the-art solutions for test smell detection and refactoring that are tailored to specific programming languages, often focusing on just one language at a time. This narrow scope severely limits the reusability of research findings across different programming environments. Consequently, researchers find themselves repeatedly building solutions from scratch to address identical problems in several languages, a process that is both complex and time-consuming. The development of more general and reusable solutions could dramatically simplify this process, but achieving this goal requires first illuminating the current landscape, identifying potential research directions, and establishing a foundation for future investigations.

To address this critical gap, our work explores the relationship between test smells and the programming languages and frameworks in which they manifest. Through a comprehensive Systematic Mapping Study (SMS), we gathered and analyzed data from papers published up to August 2025, ultimately selecting 117 papers that explicitly addressed test smells while citing specific programming languages or test frameworks. Our investigation revealed a rich landscape of 213 distinct test smells distributed across 11 programming languages (including Java, Python, JavaScript, TypeScript, C\#, and others) and 15 test frameworks (such as JUnit, PyTest, Jest, xUnit, and others). 

Our findings reveal several noteworthy patterns. Java emerges as the most frequently referenced language, with JUnit standing out as the predominant testing framework. In contrast, Python has gained substantial traction in recent years, reflecting an increasing diversity within the research landscape. Regarding test smells, \textit{Assertion Roulette} and \textit{Eager Test} are the most frequently discussed in the literature, whereas \textit{Unknown Test} demonstrates the broadest coverage across programming languages. We also identified language-specific test smells, such as those observed exclusively in TTCN-3. 

Beyond prevalence, we examined test smell criticality from two complementary perspectives: developers’ perceptions and their impact on maintainability. Interestingly, despite their frequent discussion, \textit{Assertion Roulette} and \textit{Eager Test} are not perceived as highly critical. Instead, we found the most severe test smells to be \textit{Sleepy Test}, \textit{Ignored Test}, \textit{Resource Optimism}, and \textit{Mystery Guest}. We also synthesized the refactorings, datasets, and detection/refactoring tools reported in the literature for different language and framework contexts.

Unlike previous works, our goal is neither to extend existing test smell catalogs nor to standardize test smell nomenclature. Instead, we use existing catalogs as reference sources and provide a cross-language and cross-framework synthesis of evidence reported in primary studies. In particular, unlike catalog-oriented resources such as the Open Catalog of Test Smells~\citep{Soares2023.2}, our focus is not on consolidating names and definitions, but rather on analyzing how knowledge about test smells is distributed across programming languages and testing frameworks. Likewise, unlike tool-centered mapping studies such as those by~\citet{Aljedaani2021}, we do not limit our analysis to detection tools. Tools represent only one of the six dimensions examined in this study, alongside the prevalence of test smells, their universality across languages and frameworks, associated refactorings, available datasets, and their criticality. This perspective allows us to identify which test smells generalize across ecosystems, which are language-specific, and where current research remains concentrated.

Among the key contributions of this work are:

\begin{itemize}
    \item A cross-language and cross-framework synthesis of 117 papers on test smells, organized around six research questions addressing prevalence, universality, refactorings, datasets, criticality, and tools.
    
    \item An evidence-based characterization of which test smells recur across languages and frameworks and which appear to be language-specific, including a categorization of test smells.
    
    \item A consolidated view of the evidence available in the literature for addressing test smells in different languages, including 94 refactorings for 56 test smells, 16 datasets, and 36 detection or refactoring approaches and tools.
    
    \item An evaluation of test smell criticality from two complementary perspectives: developers' perceptions and the impact of test smells on software maintainability.
    
    \item A set of insights highlighting the current concentration of research within a limited number of languages and outlining opportunities for broader and more reusable cross-language solutions.

    \item A replication package to support future extensions and validation of this research.
\end{itemize}

The remainder of this paper is organized as follows: Section \ref{sec:background} establishes the foundational concepts necessary for understanding our research, Section \ref{sec:research-methodology} details our systematic research methodology, Section \ref{sec:results-and-analysis} presents our comprehensive findings and analysis, Section \ref{sec:discussion} provides an in-depth examination and interpretation of the results, Section \ref{sec:threats-to-validity} addresses potential threats to validity and our mitigation strategies, Section \ref{sec:related-work} situates our work within the broader research landscape, and Section \ref{sec:conclusions} concludes with key insights and directions for future research.

\section{Background}\label{sec:background}

In this section, we introduce key concepts essential for understanding this work.

\subsection{Test Smells}

Code and test smells are similar concepts, denoting suboptimal practices that can lead to potential maintainability issues. While code smells refer to suboptimal coding practices in production code, test smells highlight poor engineering decisions in test code. Test smells can be particularly harmful because developers spend around one-quarter of their time engineering test code~\citep{Beller2015}. This issue becomes especially critical in environments with continuous delivery, where time to market is crucial~\citep{Tran2021}.

Although test smells do not necessarily indicate functional failures, they signal potential quality problems~\citep{Panichella2022}. Many studies have extensively examined these issues in the literature.~\citet{Deursen2001} introduced an initial catalog of test smells, describing 11 test smells and their associated refactorings. Over time, researchers have expanded this catalog to include new test smells, including those specific to certain languages \citep{Aljedaani2021}.

For readers unfamiliar with test smells, Listing \ref{lst:assertion-roulette-java-example} presents a simple example extracted from the work of~\citet{Peruma2020}. The example focuses on the test smell, \textit{Assertion Roulette}, written in Java. This smell occurs when multiple assertions appear in the same test case, but some or all assertions lack explanatory messages~\citep{Peruma2020}. This situation can become problematic because, if the test fails, developers may find it challenging to identify which assertion caused the failure~\citep{Peruma2020}.

Listing \ref{lst:assertion-roulette-python-example} shows the same example in Python. As can be seen, there are many similarities between the code snippets, especially in the structure of the assertions. The process of detecting \textit{Assertion Roulette} is the same regardless of the language. In both code examples, detecting the \textit{Assertion Roulette} test smell requires searching for multiple assertions without explanatory messages.

Listings \ref{lst:assertion-roulette-python-unittest-example} and \ref{lst:assertion-roulette-python-pytest-example} show examples of the \textit{Ignored Test} smell in Python, using Unittest and PyTest, respectively. These examples demonstrate that the similarities are even more prominent across different frameworks within the same language. The only difference between the tests is the annotation used to disable them, which generates an \textit{Ignored Test} smell.

\begin{lstlisting}[language=Java, caption=Assertion Roulette Java + JUnit example. Source: Adapted from \citet{Peruma2020}, label={lst:assertion-roulette-java-example}]
@Test
public void testCloneNonBareRepoFromLocalTestServer() throws Exception {
  Clone cloneOp = new Clone(
    false, 
    integrationGitServerURIFor("small-repo.early.git"), 
    helper().newFolder()
  );

  Repository repo = executeAndWaitFor(cloneOp);

  assertThat(
    repo, 
    hasGitObject("ba1f63e4430bff267d112b1e8afc1d6294db0ccc")
  );

  File readmeFile = new File(repo.getWorkTree(), "README");
  assertThat(readmeFile, exists());
  assertThat(readmeFile.length(), equalTo(12L));
}
\end{lstlisting}

\begin{lstlisting}[language=Python, caption=Assertion Roulette Python + PyTest example. Source: Adapted from \citet{Peruma2020}, label={lst:assertion-roulette-python-example}]
def test_clone_non_bare_repo_from_local_test_server():
    repo_uri = integration_git_server_uri_for("small-repo.early.git")
    temp_folder = new_folder()
    repo = clone_repo(repo_uri, temp_folder)

    assert "ba1f63e4430bff267d112b1e8afc1d6294db0ccc"
        in [obj.hexsha for obj in repo.git.rev_list("HEAD")]

    readme_file = temp_folder / "README"
    assert readme_file.exists()
    assert readme_file.stat().st_size == 12
\end{lstlisting}

\citet{Deursen2001} define test refactoring as transforming test code to enhance its readability or maintainability without removing test cases. A straightforward solution for the test smell shown in Listing \ref{lst:assertion-roulette-java-example} involves providing an explanatory message for each assertion~\citep{Deursen2001}. This approach is illustrated in Listing \ref{lst:refactored-assertion-roulette-java-example}.

\subsection{Test Smell Detection Tools}

Developers have created various tools to detect test smells in different programming languages, such as Java, C++, Smalltalk, and Scala, with most tools focusing on Java~\citep{Aljedaani2021}. These tools use different techniques for identifying test smells. Four notable techniques include Metrics, Rules/Heuristics, Information Retrieval, and Dynamic Tainting methodologies.

\begin{lstlisting}[language=Python, caption=Assertion Roulette Python + Unittest example. Source: Authors, label={lst:assertion-roulette-python-unittest-example}]
class PeerPriorityTest(unittest.TestCase):

    @unittest.skip("Disabled for now as this test is too flaky")
    def test_skipped_example(self):
        self.peerGroup.addConnectedEventListener(self.connectedListener)
        # ...
\end{lstlisting}

\begin{lstlisting}[language=Python, caption=Assertion Roulette Python + PyTest example. Source: Authors, label={lst:assertion-roulette-python-pytest-example}]
@pytest.mark.skip(reason="Disabled for now as this test is too flaky")
def test_skipped_example():
    self.peerGroup.addConnectedEventListener(self.connectedListener)
    # ...
\end{lstlisting}

\begin{lstlisting}[language=Java, caption=Refactored Assertion Roulette Java example. Source: Adapted from \citet{Peruma2020}, label={lst:refactored-assertion-roulette-java-example}]
@Test
public void testCloneNonBareRepoFromLocalTestServer() throws Exception {
  Clone cloneOp = new Clone(
    false, 
    integrationGitServerURIFor("small-repo.early.git"), 
    helper().newFolder()
  );

  Repository repo = executeAndWaitFor(cloneOp);
  
  // explanatory message added
  assertThat(
    "repo should has correct git object",
    repo, 
    hasGitObject("ba1f63e4430bff267d112b1e8afc1d6294db0ccc")
  );

  File readmeFile = new File(repo.getWorkTree(), "README");
  // explanatory message added
  assertThat("readme file should exist", readmeFile, exists());
  // explanatory message added
  assertThat(
    "readme file length should be 12L",
    readmeFile.length(),
    equalTo(12L)
  );
}
\end{lstlisting}

\textbf{Metrics}: This technique uses structural and semantic metrics to measure the impact of symptoms, with defined thresholds to identify smells. Tools like those developed by~\citet{VanRompaey2007} detect test smells, such as General Fixture and Eager Test, using metrics like the Number of Objects Used in setup.

\textbf{Rules/Heuristics}: These techniques identify predefined patterns in the source code, supplementing metrics. For example, developers can detect \textit{Assertion Roulette} by examining assertion statements. TSDETECT~\citep{Peruma2020}, which applies this technique to the Abstract Syntax Tree (AST), is widely used.

\textbf{Information Retrieval}: This technique extracts and normalizes information from test code, using preprocessing steps and machine learning to discern textual features. Tools like those proposed by~\citet{Lambiase2020} effectively utilize this technique.

\textbf{Dynamic Tainting}: This approach monitors code during execution, analyzing runtime data with predefined taint values. \citet{Zhang2014} used this method in their work.

Developers commonly use rule/heuristic-based detection due to the presence of well-defined rules, while they use metrics-based detection less frequently due to the limitations of metrics. They typically apply dynamic-based detection to identify test dependencies and rotten green tests. Information retrieval techniques are also helpful, though they may struggle without relevant features.

\citet{Aljedaani2021} identified 22 test smell detection tools in the literature, which together support 66 test smells. Among the detection tools, JNose and TSDetect are the ones that support the largest number of test smells, with 21 and 19, respectively~\citep{Aljedaani2021}. The sets of test smells supported by these tools are largely similar, differing only in the inclusion of \textit{Verbose Test}, \textit{Ignored Test}, \textit{For Testers Only}, and \textit{Assertionless} smells. On the list of tools reported by~\citet{Aljedaani2021}, two can be pointed out that support refactoring: TestHound and TRex.

\section{Research Methodology}\label{sec:research-methodology}

This section outlines the procedures used to conduct a systematic mapping study on test smells across programming languages. The mapping process follows the guidelines for systematic mapping studies in software engineering~\citep{Petersen2008, Petersen2015}.

\subsection{Goal And Research Questions}

The main goal of this work is to address the gap identified in Section \ref{sec:related-work} by analyzing test smells in the literature from a programming language perspective, providing comparisons and insights on this topic. To achieve this goal, we defined the following research questions:

\textbf{RQ1.} \textit{Which programming languages and frameworks are most frequently addressed in discussions of test smells, and what are the most prevalent test smells associated with each?} This question aims to identify the prevalence of certain programming languages and frameworks in test smell discussions and determine the most frequently addressed test smells for each.

\textbf{RQ2.} \textit{What are the universal test smells across all programming languages and frameworks, and which test smells are specific to particular languages?} This question seeks to identify test smells common across all (or most) programming languages and frameworks, as well as those specific to certain languages and frameworks.

\textbf{RQ3.} \textit{What refactorings are commonly recommended for addressing test smells in each programming language and framework?} This question aims to identify the techniques used to refactor test smells across different languages and frameworks.

\textbf{RQ4.} \textit{What datasets of test smells are available for each programming language and framework?} This question aims to identify and catalog datasets of test smells across different languages and frameworks.

\textbf{RQ5.} \textit{Which test smells are most critical across different programming languages and frameworks?} This question aims to gather evidence from the literature about the severity of various test smells in different programming languages and frameworks. We analyzed criticality from two perspectives: a general developer's perception of the test smell and the impact of the test smell on maintainability.

\textbf{RQ6.} \textit{What test smell detection or refactoring tools have been proposed for each programming language and framework?} This question seeks to identify detection and refactoring tools proposed for each language and framework, providing practical and useful information for both academics and practitioners aiming to address test smells in specific languages or frameworks.

With these research questions, we aimed not only to provide a language perspective on test smells through RQ1 and RQ2, but also to offer insights that can support future research. These insights can help develop solutions to expand the set of studied languages in the context of test smells or create general approaches for detecting and refactoring test smells across multiple languages. Specifically, our contributions include common refactorings for test smells that could be implemented in new tools (RQ3), datasets of test smells in various languages that could be used to validate new studies (RQ4), identification of the most critical test smells that require greater attention (RQ5), and a list of tools that can serve as a reference or comparison basis for new studies (RQ6).

\subsection{Search Strategy}


From the research questions, we identified relevant terms and their synonyms to use as keywords for the literature search. The keywords were organized into three groups, connected by AND clauses. The search string employed was:

\textit{(``test smell'' OR ``test smells'') AND (``tool'' OR ``detect'' OR ``refactoring'' OR ``dataset'') AND (``language'' OR ``languages'' OR ``code'')}.

The first group (``test smell'' OR ``test smells'') captures the core concept of the study. The second group (``tool'' OR ``detect'' OR ``refactoring'' OR ``dataset'') aligns with the scope of supporting evidence we aimed to synthesize, covering detection and refactoring tools (RQ6), refactorings (RQ3), and datasets (RQ4). The third group (``language'' OR ``languages'' OR ``code'') restricts the results to works that explicitly address programming languages or code, which is central to RQ1 and RQ2 and to our overall goal of analyzing test smells from a programming language perspective. All three AND clauses were retained to ensure that the search results were aligned with our research questions while keeping the screening load manageable. The search string was validated through pilot testing with well-known reference works and iterative refinement across all digital libraries.

To assess whether the search string was overly restrictive, we performed a sensitivity check in the ACM Digital Library using the query \textit{(``test smell'' OR ``test smells'')}. The full search string returned only four fewer results than this broader query. This small difference indicates that the additional keyword groups do not substantially reduce the result set. We also examined the first 20 results from the broader query and found that 8 of these did not explicitly mention a programming language in the title or abstract. Including the third group (``language'' OR ``languages'' OR ``code'') therefore helps to focus the results on papers that address the language and framework dimension, which is central to our research questions, without excluding a substantial body of relevant work.

To capture as many relevant works as possible, we left the start of the search period open and limited only the end to August 2025. We also selected relevant and high-quality digital libraries to conduct the automatic search, as shown in Table \ref{tab:digital-libraries}.

\begin{table}[h]
    \scriptsize
    \centering
    \caption{Digital libraries for automatic search}
    \label{tab:digital-libraries}
    \begin{tabular}{ll}
        \toprule
        \textbf{Name} & \textbf{URL} \\
        \midrule
        ACM Digital Library & \texttt{http://portal.acm.org} \\
        El Compendex        & \texttt{http://www.engineeringvillage.com} \\
        IEEE Digital Library & \texttt{http://ieeexplore.ieee.org} \\
        Science@Direct      & \texttt{http://www.sciencedirect.com} \\
        Scopus              & \texttt{http://www.scopus.com} \\
        Springer Link       & \texttt{http://link.springer.com} \\
        \bottomrule
    \end{tabular}
\end{table}

\subsection{Selection Strategy}

The search described in the previous sections yielded a substantial number of studies (787 in total). The first step was to remove duplicate papers, which resulted in the elimination of 333 duplicates. Next, we applied the inclusion and exclusion criteria detailed in Table \ref{tab:inclusion-exclusion-criteria}.

\begin{table}[h]
    \scriptsize
    \centering
    \caption{Inclusion and Exclusion Criteria}
    \label{tab:inclusion-exclusion-criteria}
    \begin{tabular}{p{6.5cm} p{6.5cm}}
        \toprule
        \textbf{Inclusion Criteria} & \textbf{Exclusion Criteria} \\
        \midrule
        Addresses test smells in one or more programming languages or testing frameworks & Does not address test smells, or does not address them in the context of programming languages or testing frameworks \\
        Papers written in English & Papers written in languages other than English \\
        Peer-reviewed journal articles, full conference and workshop papers, and short papers or extended abstracts reporting conceptual or empirical results related to test smells & Secondary or tertiary studies, posters, tutorials, keynotes, theses, books without primary-study content, and non-scientific sources \\
        & Duplicate papers \\
        & Records that only report the proceedings of an event (e.g., volumes or web pages listing the papers of a conference or workshop, tables of contents, or editorial prefaces) without presenting an individual primary study \\
        \bottomrule
    \end{tabular}
\end{table}

We included peer-reviewed journal articles, full conference and workshop papers, as well as short papers and extended abstracts, since relevant contributions on test smells are sometimes published in these formats. Excluding them could therefore omit important evidence. We only excluded records that referred to the proceedings themselves (such as tables of contents or editorial prefaces) and did not present an individual primary study.

\begin{figure*}[!h]
    \centering
    \includegraphics[width=\textwidth]{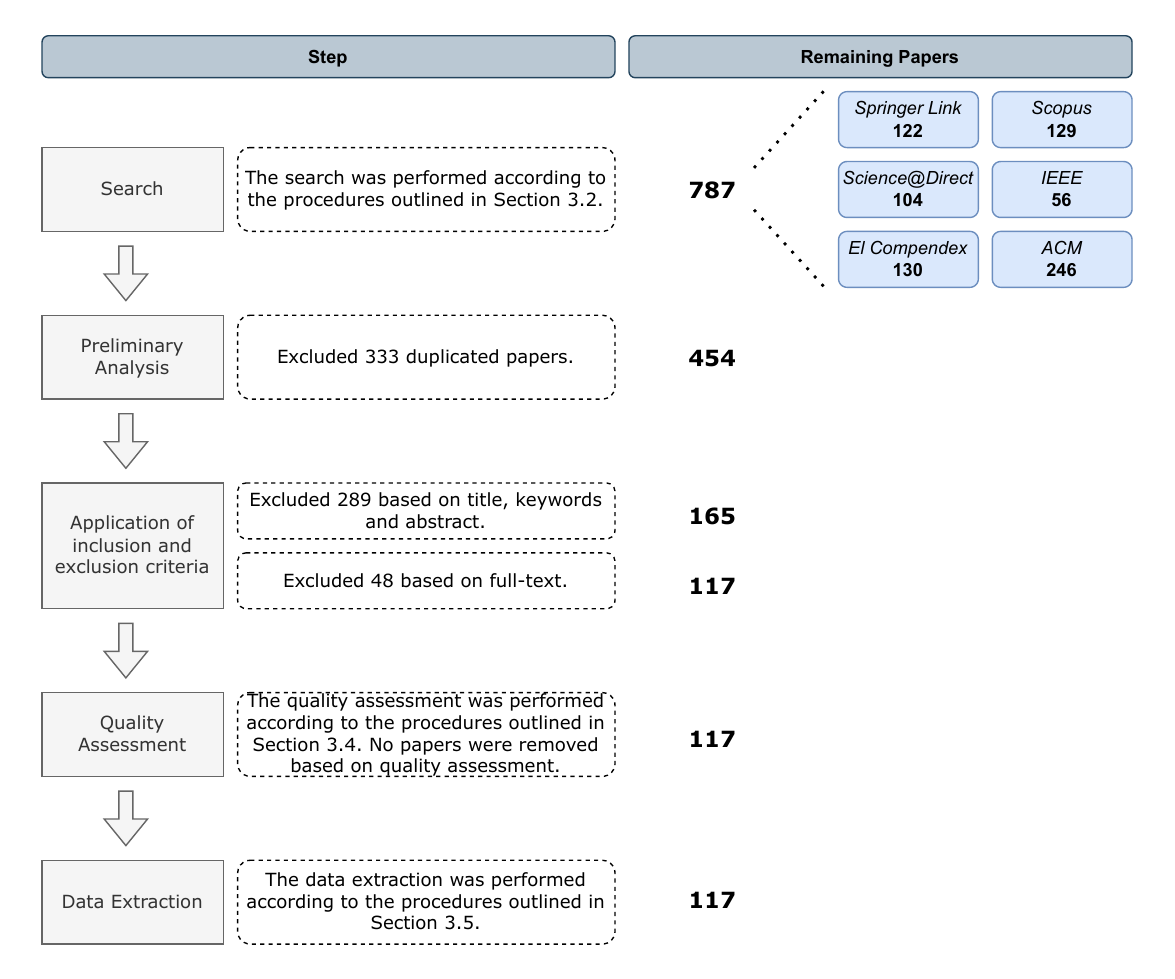}
    \caption{Selection Process.}
    \label{fig:selection-strategy-process}
\end{figure*}

By applying the inclusion and exclusion criteria, 289 papers were excluded based solely on their titles, keywords, and abstracts. Following this, we reviewed the full texts of the remaining papers. During this stage, we removed 48 papers based on the inclusion and exclusion criteria applied to the full texts. This process resulted in a final set of 117 papers. Subsequently, we conducted a quality assessment of these papers, but no papers were excluded based on this assessment. The entire process is illustrated in Figure \ref{fig:selection-strategy-process}.

For the paper selection process, we employed a peer review approach involving three of the four co-authors. The review team consisted of two doctoral-level researchers with extensive expertise in test smells and software testing, and one master's student with moderate experience in the domain. This composition ensured both methodological rigor and diverse perspectives during the selection process. We conducted the selection process in two distinct phases: (1) initial screening based on titles, abstracts, and keywords, and (2) full-text review for papers that passed the initial screening. In practice, the reviewers' initial assessments were usually aligned. To maintain methodological transparency, any papers with discordant initial assessments were automatically advanced to the full-text review phase, allowing for more comprehensive evaluation. This approach ensured we did not prematurely exclude potentially relevant papers due to limited information in abstracts or titles. When reviewers disagreed on study inclusion, they discussed their perspectives and revisited the inclusion and exclusion criteria until reaching consensus. In cases of disagreement at the title or abstract level, papers were advanced to full-text screening rather than excluded, following the adaptive reading-depth recommendation for mapping studies when abstract-level information is insufficient~\citep{Petersen2008}.

\subsection{Quality Assessment}


We conducted a quality assessment to evaluate the strength of the evidence in the studies included in this SMS. The assessment questions (Table~\ref{tab:quality-assessment-form}) were derived from the 11 quality criteria proposed by~\citet{machado2014}, which cover four key aspects: reporting, rigor, credibility, and relevance. Following established guidelines for systematic reviews and mapping studies in software engineering, we adapted these criteria to the specific goals of our SMS on test smells. In particular, the questions were designed to capture: (i) the clarity and completeness of the study description, (ii) the soundness of the empirical design and evaluation procedures, (iii) the transparency of reporting (including the description of datasets, tools, and refactoring strategies), and (iv) the practical and research relevance of the reported findings.

To evaluate the reliability of the quality assessment, we applied the checklist to a random subset of primary studies. Each question was rated using the same binary scale later adopted in the full assessment (``yes''~$=$~1 and ``no''~$=$~0). Disagreements were discussed until consensus was reached, which helped calibrate the interpretation of each question among the reviewers.

For the final assessment, each question was scored using a binary scale: ``yes'' (1) or ``no'' (0). With 11 questions, the maximum possible score was 11. The studies obtained scores ranging from 5 to 10, with a median of 9 and an average of 8.61. These results indicate that the papers included in this SMS generally exhibit high methodological quality. No studies were excluded based on the quality assessment; instead, the scores were used to contextualize our findings and to highlight cases in which conclusions rely primarily on studies with comparatively lower quality scores.

\begin{table}[h]
    \scriptsize
    \centering
    \caption{Quality Assessment Questions}
    \label{tab:quality-assessment-form}
    \begin{tabular}{p{2cm} p{9cm} p{2cm}}
        \toprule
        \textbf{Number} & \textbf{Question} & \textbf{Issue} \\
        \midrule
        1 & Is the paper based on research and not merely a ‘‘lessons learned’’ report based on expert opinion? & Reporting \\
        2 & Is there a clear statement of the aims of the research? & Reporting \\
        3 & Is there an adequate description of the context in which the research was carried out? & Reporting \\
        4 & Was the research design appropriate to address the aim of the research? & Rigour \\
        5 & Was there a control group with which to compare the treatments? & Rigour \\
        6 & Was the data collected in a way that addressed the research issue? & Rigour \\
        7 & Was the data analysis sufficiently rigorous? & Rigour \\
        8 & Has the relationship between researcher and participants been considered to an adequate degree? & Credibility \\
        9 & Is there a clear statement of findings? & Credibility \\
        10 & Is the study valuable for research or practice? & Relevance \\
        11 & Are there any practitioner-based guidelines? & Relevance \\
        \bottomrule
    \end{tabular}
\end{table}

\subsection{Data Extraction and Synthesis}

For data extraction, we designed a form, shown in Table \ref{tab:data-extraction-form}, to address the six research questions defined for this SMS. The questions in the form cover the following aspects: test smells, programming languages and frameworks mentioned in the studies, detection or refactoring tools for test smells, and evidence regarding the criticality of each test smell. We extracted all data in plain text format.


Some test smells appear in the literature under different names, even though they describe the same underlying issue. The open catalog by \citet{Soares2023.2} already reflects this situation, as several catalog entries include multiple alternative names associated with the same scent. For example, the catalog treats \textit{Complex Test} and \textit{Verbose Test} as alternative labels for the same problem, namely \textit{Obscure Test}. To avoid reporting duplicate test smells, we relied on this open catalog as a canonical reference for terminology normalization during data extraction. In other words, we used the catalog to standardize the names and definitions adopted when synthesizing evidence from the primary studies.

When the catalog did not explicitly indicate that two smell names referred to the same issue, or when we judged that a study's smell definition diverged from all catalog entries, we manually analyzed the definitions and examples (from the catalog when available and from the original paper otherwise). We then compared them with the smells already identified in our analysis to determine whether they represented duplicates. One such example is \textit{Redundant Print}, which in some papers is referred to as \textit{Console Printing}, although the definitions describe the same underlying issue. All cases of this kind are documented in our replication package, where we report the standardized name we adopted and the alternative names identified in the primary studies.

To analyze how test smells are distributed across languages within each category, we relied on the same top-level taxonomy used in the open catalog (code-related, dependencies, design-related, issues in test steps, test execution - behavior, and test semantic - logic), without introducing ad hoc categories. For each normalized smell present in the open catalog, we directly adopted its assigned category. For smells not listed in the open catalog, we manually assigned the most appropriate category by comparing each smell’s definition from the primary studies with the catalog’s category descriptions and with similar or related smells already documented. We recorded a brief justification for each assignment, and any disagreements were discussed among the authors until consensus was reached. Table~\ref{tab:outside-catalog-categories} (\ref{ap:outside-catalog-categories}) lists these non-catalog smells along with their assigned categories. The corresponding justifications are available in the replication package.

For RQ3, we manually reviewed the primary studies and recorded refactoring techniques only when the paper explicitly named a refactoring or clearly stated that a given technique, operation, or change could remove, mitigate, or refactor a specific test smell. We did not infer new refactoring labels from code examples or from general discussions when the study did not make this relationship explicit. To reduce bias in this process, we consistently applied the same decision criteria across all studies.

\begin{table}[h]
    \scriptsize
    \centering
    \caption{Data Extraction Form}
    \label{tab:data-extraction-form}
    \begin{tabular}{p{3cm} p{8cm} p{2cm}}
        \toprule
        \textbf{Question} & \textbf{Description} & \textbf{RQ} \\
        \midrule
        Languages and Frameworks Mentioned & Programming languages and frameworks discussed in the study & RQ1/RQ2 \\
        Test Smells Identified & Test smells associated with the mentioned languages/frameworks & RQ1/RQ2 \\
        Test Smells Refactorings & Refactorings for the identified test smells & RQ3 \\
        Test Smells Datasets & Test smells datasets used or referenced on the selected papers & RQ4 \\
        Test Smells Criticality & Assessment or discussion of the severity or impact of the identified test smells & RQ5 \\
        Detection or Refactoring Tool Proposed & Description of any detection or refactoring tools proposed in the study & RQ6 \\
        \bottomrule
    \end{tabular}
\end{table}

\section{Results}\label{sec:results-and-analysis}

In this section we present an overview of the results we obtained and answer the research questions.

\subsection{Overview}

\begin{figure*}[!h]
    \centering
    \includegraphics[width=\textwidth]{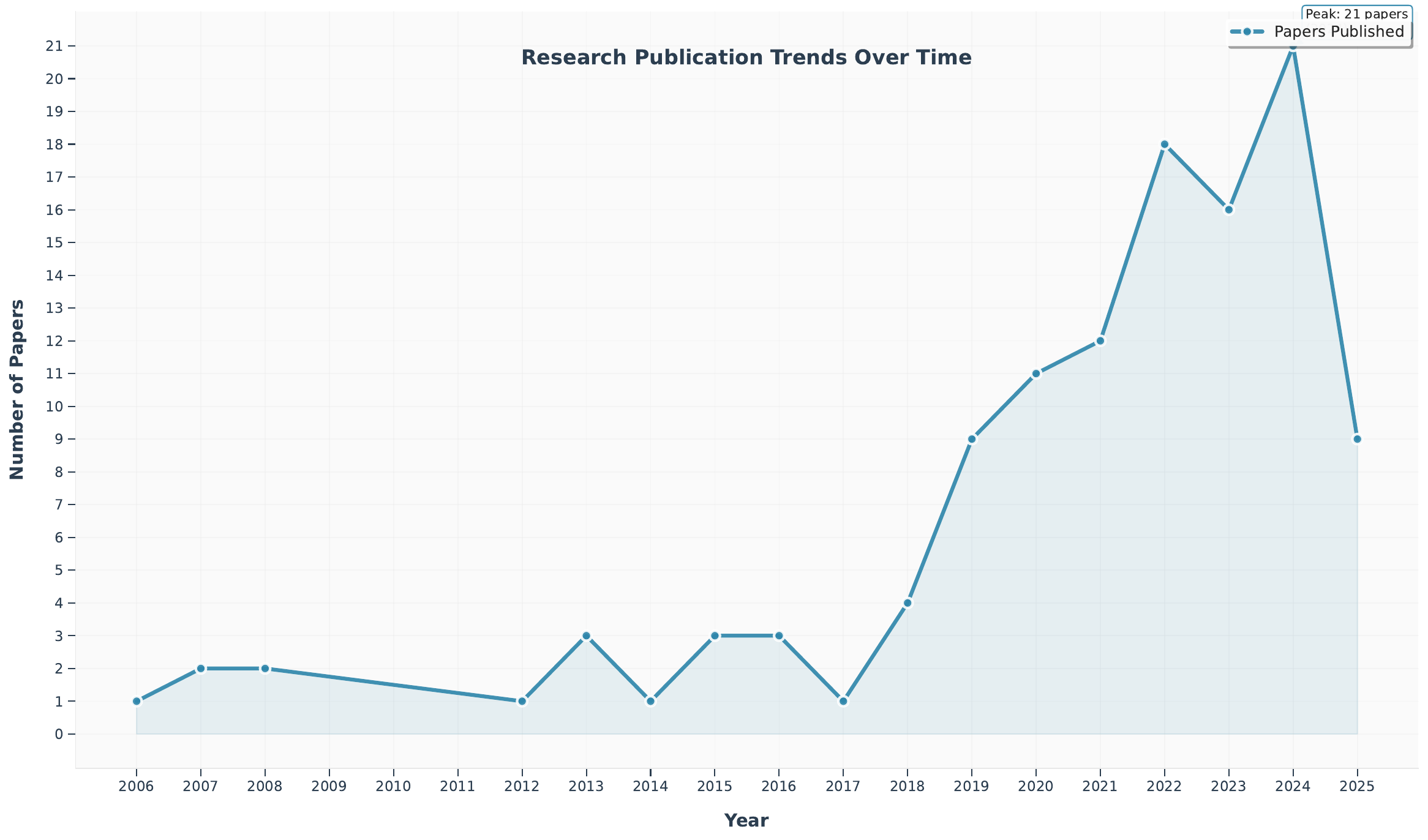}
    \caption{Final Papers Per Year.}
    \label{fig:final-papers-per-year}
\end{figure*}

After applying the process shown in Figure \ref{fig:selection-strategy-process}, we retained 117 papers. Figure \ref{fig:final-papers-per-year} indicates the number of studies (out of the 117 accepted) per year. The figure demonstrates a notable increase in studies involving test smells since 2018, indicating that researchers have shown growing interest in this topic.

Our analysis revealed several key quantitative insights: (1) The 117 papers collectively identified 213 distinct test smells across 11 programming languages and 15 testing frameworks, with Java/JUnit being the most frequently studied combination (appearing in 62 papers); (2) The most universal test smell was \textit{Unknown Test}, appearing in 11 different language/framework combinations, followed by \textit{Conditional Test Logic}, \textit{Exception Handling}, and \textit{Magic Number}, each appearing in 10 combinations; (3) A total of 94 distinct refactoring techniques were identified, with \textit{Extract Method} being the most frequently mentioned (44 times); (4) The criticality analysis covered 33 test smells, with \textit{Sleepy Test} identified as the most critical; and (5) The tooling landscape includes 36 tools, with 61\% currently available and only 14\% supporting both detection and refactoring capabilities.

\subsection{RQ1: Which programming languages and frameworks are most frequently addressed in discussions of test smells, and what are the most prevalent test smells associated with each?}

To answer this research question, we analyzed all the languages and frameworks related to test smells that the selected papers mentioned. The analyzed studies mention a total of 11 programming languages. The papers also mention 15 different testing frameworks: 1 for C\#, 3 for Java, 2 for JavaScript, 7 for Python, 1 for Scala, 2 for TypeScript, and one generic framework, which is Robot~\citep{bisht2013}. The complete language/framework frequency table is provided in~\ref{ap:rq1-languages-frameworks} (Table~\ref{tab:languages-frameworks}).

The languages identified span different typing systems and programming paradigms. Static typing languages include C, C\#, C++, Java, Kotlin, Scala, and TTCN-3, while dynamic typing languages include JavaScript, Python, and Smalltalk. The majority of languages (8 out of 11) use strong typing, with only C and C++ using weak typing. Most languages support multiple paradigms, with Java, C\#, C++, JavaScript, Kotlin, Python, and TypeScript being multi-paradigm languages supporting object-oriented, functional, and/or procedural programming. Only Smalltalk is purely object-oriented, while TTCN-3 is declarative and specifically designed for telecommunications testing~\citep{grabowski2003}.

\begin{figure*}[!h]
    \centering
    \includegraphics[width=\textwidth]{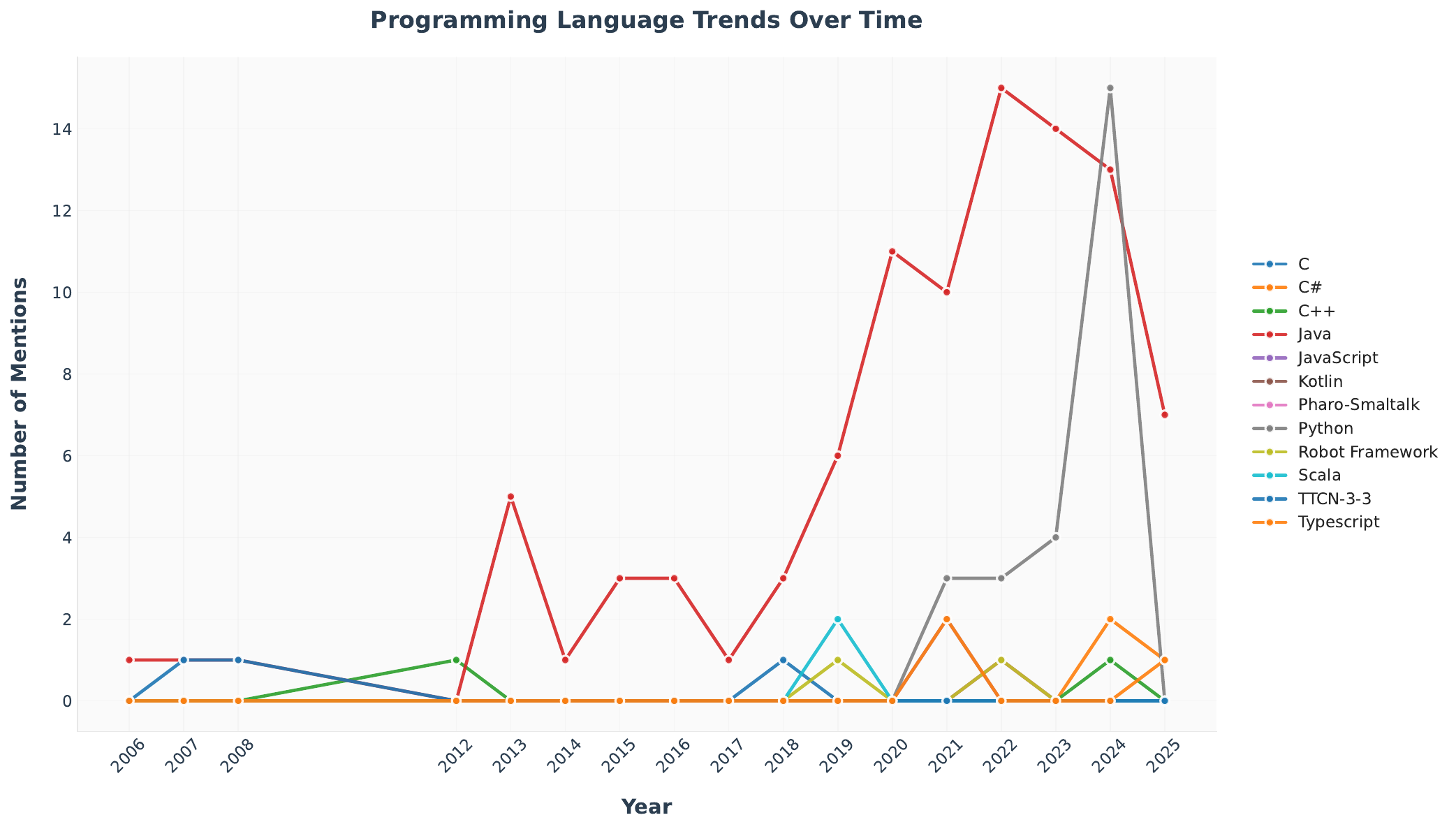}
    \caption{Language Trend.}
    \label{fig:language-trend}
\end{figure*}

Figure \ref{fig:language-trend} shows a trend of how often each language presented in Table \ref{tab:languages-frameworks} has been mentioned over the years. As can be seen in the figure, Java appeared almost every year, and in the most recent years, the number of mentions has increased. Python is a language that has started to appear with some frequency in the last few years. Some languages, like C and C++, have only been cited in one year. This trend shows that the language most frequently mentioned in the context of test smells is Java. Python, on the other hand, has only begun to appear regularly in the last few years, but its frequency of mentions has grown rapidly, surpassing Java in 2024. Other languages lack analysis in the context of test smells.

\begin{figure*}[!h]
    \centering
    \includegraphics[width=\textwidth]{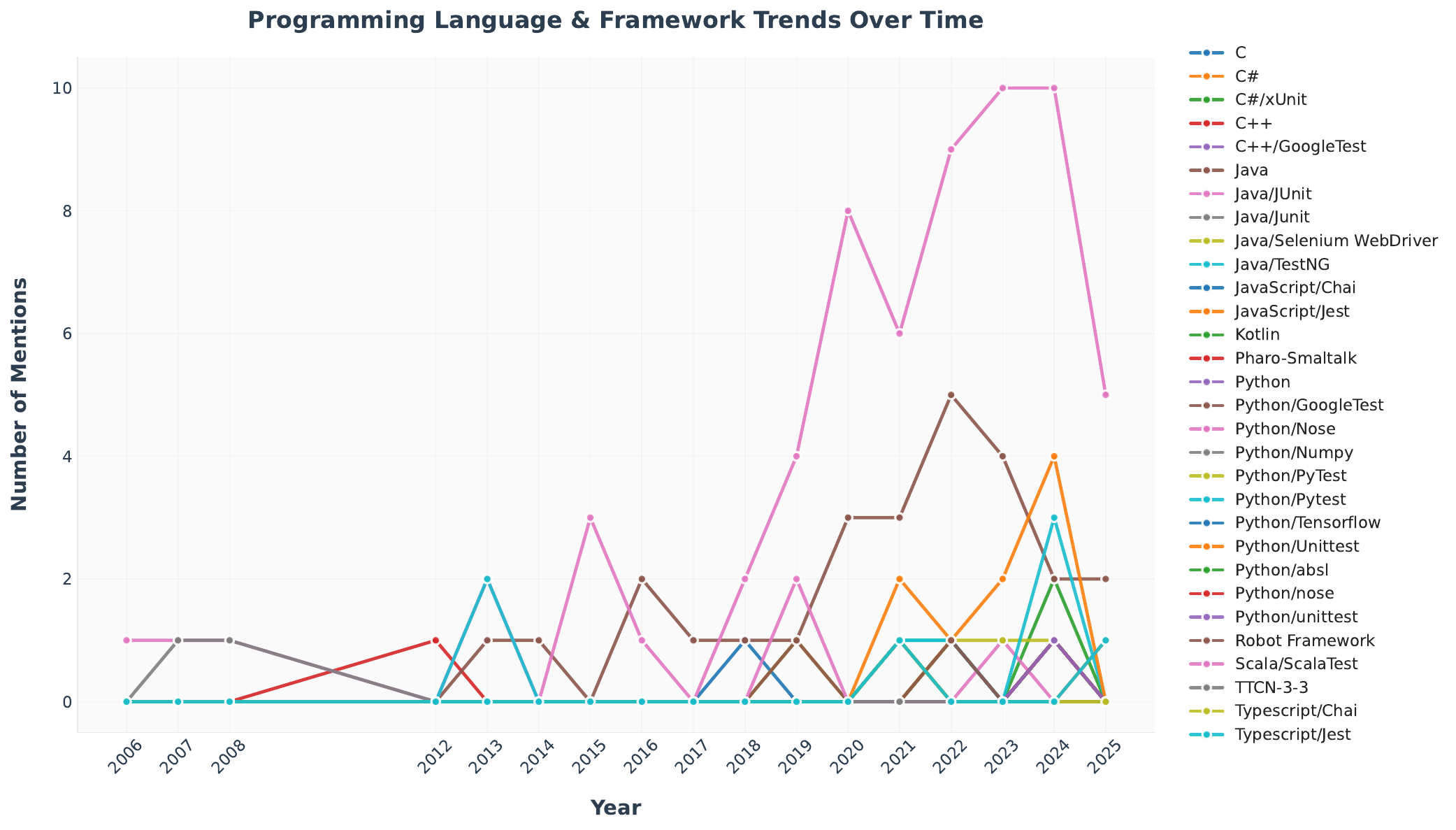}
    \caption{Language/Framework Trend.}
    \label{fig:language-framework-trend}
\end{figure*}

Figure \ref{fig:language-framework-trend} shows the same trend for the frameworks associated with each language. It demonstrates that the framework most commonly associated with Java in the majority of the papers is JUnit.

We analyzed the test smells most frequently discussed in relation to tests in each language. The test smell most commonly addressed in discussions about test smells in Java is \textit{Eager Test}, with 51 mentions. In Python, the most widely discussed test smells are \textit{Conditional Test Logic} and \textit{Exception Handling}, with ten mentions each. The other languages have few mentions of each test smell.

We also examined the most frequently addressed test smells across all languages. \ref{ap:rq1-top-smells} (Table~\ref{tab:top-20-test-smells}) shows the top 20 most mentioned test smells across all languages. The table shows that the test smells \textit{Assertion Roulette} and \textit{Eager Test} stand out as the most cited. In this table, we omitted languages that do not mention any of these 20 test smells. One interesting fact is that Java is related to all 20 test smells, while Python is associated with 19 of them.

We also analyzed the distribution of test smells across programming languages by category. For this, we classified the test smells according to the types defined in the test smell catalog of \citet{Soares2023.2}, which are: code-related, dependencies, design-related, issues in test steps, test execution (behavior), and test semantics (logic). Table~\ref{tab:test-smells-per-category} shows the distribution of test smells across the languages. 


The most common types of test smells are code-related, followed by test semantic (logic) issues and issues in test steps. The latter two occur at very similar frequencies. The remaining categories appear less frequently overall. Java exhibits a relatively balanced distribution of test smell types, with no single category dominating as strongly as in some more specialized languages. In Python, issues in test steps represent the largest share, followed closely by code-related test smells. Together, these two categories account for a substantial portion of the occurrences reported for the language. An interesting case is TTCN-3, where code-related issues clearly dominate. This behavior may be associated with the fact that TTCN-3 is a language specifically designed for writing test code; therefore, most problems tend to be code-centric rather than belonging to other categories, such as dependencies or design-related issues. Another notable observation is that dynamically typed languages (such as Python, TypeScript, which supports optional typing, and JavaScript) exhibit higher proportions of test execution (behavioral) smells compared to Java when considering within-language distributions.

\begin{table}[!ht]
    \scriptsize
    \centering
    \caption{Test Smell Categories Per Language - Counts and Percentages}
    \label{tab:test-smells-per-category}
    \begin{tabular}{c|cccccc}
    \toprule
        \textbf{Language}
        & \textbf{Code} & \textbf{Dep} & \textbf{Design} & \textbf{Steps} & \textbf{Exec} & \textbf{Sem} \\
    \midrule
        \multirow{2}{*}{C}
        & 1 & 0 & 0 & 0 & 0 & 0 \\
        & 100.0\% & 0.0\% & 0.0\% & 0.0\% & 0.0\% & 0.0\% \\
        \midrule
        \multirow{2}{*}{C\#}
        & 8 & 1 & 4 & 6 & 2 & 6 \\
        & 29.6\% & 3.7\% & 14.8\% & 22.2\% & 7.4\% & 22.2\% \\
        \midrule
        \multirow{2}{*}{C++}
        & 5 & 0 & 1 & 2 & 0 & 2 \\
        & 50.0\% & 0.0\% & 10.0\% & 20.0\% & 0.0\% & 20.0\% \\
        \midrule
        \multirow{2}{*}{Java}
        & 209 & 94 & 46 & 144 & 45 & 171 \\
        & 29.5\% & 13.3\% & 6.5\% & 20.3\% & 6.3\% & 24.1\% \\
        \midrule
        \multirow{2}{*}{Kotlin}
        & 0 & 0 & 0 & 1 & 0 & 0 \\
        & 0.0\% & 0.0\% & 0.0\% & 100.0\% & 0.0\% & 0.0\% \\
        \midrule
        \multirow{2}{*}{Python}
        & 36 & 9 & 15 & 38 & 16 & 20 \\
        & 26.9\% & 6.7\% & 11.2\% & 28.4\% & 11.9\% & 14.9\% \\
        \midrule
        \multirow{2}{*}{Scala}
        & 0 & 2 & 0 & 4 & 0 & 6 \\
        & 0.0\% & 16.7\% & 0.0\% & 33.3\% & 0.0\% & 50.0\% \\
        \midrule
        \multirow{2}{*}{Smalltalk}
        & 1 & 0 & 0 & 0 & 0 & 0 \\
        & 100.0\% & 0.0\% & 0.0\% & 0.0\% & 0.0\% & 0.0\% \\
        \midrule
        \multirow{2}{*}{TTCN-3}
        & 56 & 0 & 6 & 6 & 0 & 6 \\
        & 75.7\% & 0.0\% & 8.1\% & 8.1\% & 0.0\% & 8.1\% \\
        \midrule
        \multirow{2}{*}{Typescript}
        & 4 & 2 & 1 & 2 & 4 & 5 \\
        & 22.2\% & 11.1\% & 5.6\% & 11.1\% & 22.2\% & 27.8\% \\
        \midrule
        \multirow{2}{*}{Javascript}
        & 3 & 2 & 1 & 2 & 2 & 4 \\
        & 21.4\% & 14.3\% & 7.1\% & 14.3\% & 14.3\% & 28.6\% \\
        \midrule
        \multirow{2}{*}{Robot Framework}
        & 16 & 4 & 2 & 12 & 1 & 6 \\
        & 39.0\% & 9.8\% & 4.9\% & 29.3\% & 2.4\% & 14.6\% \\
    \bottomrule
    \end{tabular}

    \begin{tablenotes}
    \small
    \item \textbf{Abbreviations:} Code = Code related, Dep = Dependencies, Design = Design related, Steps = Issues in test steps, Exec = Test execution (behavior), Sem = Test semantic (logic)
    \end{tablenotes}
\end{table}

\rqtakeaway{RQ1 reveals a strong concentration of evidence surrounding Java and JUnit, whereas Python has gained visibility only in recent years. Overall, the cross-language landscape remains uneven, with detailed frequency distributions provided in \ref{ap:rq1-languages-frameworks} and~\ref{ap:rq1-top-smells}.}

\subsection{RQ2: What are the universal test smells across all programming languages and frameworks, and which test smells are specific to particular languages?}

To address this question, we analyzed test smells that appeared in multiple languages or frameworks, as well as those that appeared in only a single language or framework. We defined universal test smells as those present in at least two languages, indicating that there is evidence in the literature that these test smells can occur across multiple languages. Conversely, language-specific test smells are those observed in only one language, indicating that, to date, there is no evidence in the current literature that these test smells occur in multiple languages.

Table \ref{tab:universal-test-smells} shows the test smells that appeared in multiple languages or frameworks. In total, 37 test smells appeared in more than one language or framework, with mentions ranging from 2 to 11 across different languages or frameworks. As seen in the table, the test smell that appeared in the most languages or frameworks was \textit{Unknown Test}, with 11 different language/framework associations, followed by \textit{Conditional Test Logic}, \textit{Exception Handling}, and \textit{Magic Number}, which were each associated with 10 different languages/frameworks. \ref{ap:rq1-top-smells} (Table \ref{tab:top-20-test-smells}) shows that these test smells are also among the most frequently cited.

\begin{scriptsize}
\begin{table}[!ht]
    \scriptsize
    \centering
    \caption{Test Smells That Appear in More Than One Language or Framework}
    \label{tab:universal-test-smells}
    \begin{tabular}{lc}
    \toprule
    \textbf{Test Smell} & \textbf{N. of Languages/Frameworks} \\
    \midrule
    \textit{Unknown Test} & 11 \\
    \textit{Conditional Test Logic} & 10 \\
    \textit{Exception Handling} & 10 \\
    \textit{Magic Number} & 10 \\
    \textit{Assertion Roulette} & 9 \\
    \textit{Duplicate Assert} & 9 \\
    \textit{Eager Test} & 9 \\
    \textit{Redundant Print} & 9 \\
    \textit{Ignored Test} & 8 \\
    \textit{Sleepy Test} & 8 \\
    \textit{Obscure Test} & 7 \\
    \textit{Redundant Assertion} & 7 \\
    \textit{Empty Test} & 6 \\
    \textit{General Fixture} & 6 \\
    \textit{Lazy Test} & 6 \\
    \textit{Mystery Guest} & 6 \\
    \textit{Resource Optimism} & 5 \\
    \textit{Sensitive Equality} & 5 \\
    \textit{Constructor Initialization} & 4 \\
    \textit{Duplicated Test Code} & 4 \\
    \textit{Obscure In-Line Setup} & 4 \\
    \textit{Dead Field} & 3 \\
    \textit{Default Test} & 3 \\
    \textit{Dependent Test} & 3 \\
    \textit{Lack Of Cohesion Of Test Methods} & 3 \\
    \textit{Non-Functional Statement} & 3 \\
    \textit{Programming Paradigms Blend} & 3 \\
    \textit{Redundant Print} & 3 \\
    \textit{Test Maverick} & 3 \\
    \textit{Undefined Test} & 3 \\
    \textit{Verifying In Setup Method} & 3 \\
    \textit{Bad Naming} & 2 \\
    \textit{Inappropriate Assertion} & 2 \\
    \textit{Indirect Testing} & 2 \\
    \textit{Lack Of Cohesion} & 2 \\
    \textit{Private Method Test} & 2 \\
    \textit{Redundant Test} & 2 \\
    \bottomrule
    \end{tabular}
\end{table}
\end{scriptsize}

\ref{ap:specific-test-smells} shows the list of test smells that appeared in a single language or framework. In total, we identified 176 test smells in the languages or frameworks: C++, Java (without specifying any framework), Java/JUnit, Java/TestNG, Python/PyTest, Python/Unittest, Robot Framework, Smaltalk, TTCN-3, and Typescript/Jest. As shown in the table, Java is the language with the most test smells that appear exclusively within a single language, which aligns with the fact that Java (and its frameworks) was the most frequently mentioned in papers addressing test smells. An interesting observation is that, apart from Java, the languages or frameworks with the highest number of test smells exclusive to a single language were those specifically designed for writing tests, such as Robot~\citep{bisht2013} and TTCN-3~\citep{grabowski2003}.

\rqtakeaway{RQ2 indicates that a relatively small group of test smells recurs across multiple language and framework contexts, whereas a much larger set is reported in only a single context. This pattern suggests that some smells are clearly cross-language, while the generalizability of many others remains uncertain in the current literature.}

\subsection{RQ3: What refactorings are commonly recommended for addressing test smells in each programming language and framework?}


To answer this question, we considered only the refactoring techniques explicitly reported by the reviewed studies for specific test smells, following the extraction rule described in Section~\ref{sec:research-methodology}. Using this procedure, we identified 56 test smells with associated refactorings, resulting in 94 distinct refactorings across three programming languages (Java, Scala, and Python) and seven testing frameworks (JUnit, TestNG, ScalaTest, Robot Framework, Unittest, PyTest, and Nose). Some test smells had multiple refactorings reported. For example, for the \textit{Assertion Roulette} smell, two possible refactorings are \textit{Add Assertion Explanation} (demonstrated in Listing \ref{lst:refactored-assertion-roulette-java-example}) and \textit{Extract Method} (illustrated in Listing \ref{lst:refactored-assertion-roulette-java-example-extract-method}). Additionally, some refactorings address multiple test smells; for instance, \textit{Extract Method} is also used to refactor the \textit{Duplicate Assert} smell.

The most frequently mentioned refactoring technique was \textit{Extract Method}, appearing 44 times across different test smells, followed by \textit{Add Assertion Explanation} (16 mentions) and \textit{Setup External Resource} (9 mentions). Other commonly cited refactorings include \textit{Extract Class} (8 mentions), \textit{Use Parameterized Test} (7 mentions), and \textit{Introduce Equality Method} (6 mentions). The test smell with the most diverse refactoring options was \textit{Exception Handling}, with 11 different refactoring techniques reported, followed by \textit{Assertion Roulette} and \textit{Conditional Test Logic}, each with seven different refactoring approaches.

Table \ref{tab:test-smells-favorite-refactoring} presents the most frequently cited refactorings. We omitted from this table test smells with only a single reported refactoring, as well as those for which all refactorings have the same number of mentions. As shown, the most common refactorings for the test smells \textit{Assertion Roulette} and \textit{Eager Test} (the two most frequently cited test smells) are \textit{Add Assertion Explanation} and \textit{Extract Method}, respectively. Additionally, the \textit{Extract Method} appears most often overall, as it can address multiple test smells. A comprehensive list of refactorings, organized by language and framework, is available in \ref{ap:test-smells-refactorings}.

\begin{table}[!ht]
    \centering
    \caption{Test Smells and Favorite Refactorings}
    \label{tab:test-smells-favorite-refactoring}
    \begin{scriptsize}
    \begin{tabular}{llc}
        \toprule
        \textbf{Test Smell} & \textbf{Refactoring} & \textbf{N. of Mentions} \\
        \midrule
        \textit{Assertion Roulette} & Add assertion explanation & 16 \\
        \textit{Eager Test} & Extract method & 11 \\
        \textit{Duplicate Assert} & Extract method & 7 \\
        \textit{Sensitive Equality} & Introduce equality method & 6 \\
        \textit{Duplicated Test Code} & Extract method & 5 \\
        \textit{Mystery Guest} & Setup external resource & 5 \\
        \textit{Resource Optimism} & Setup external resource & 4 \\
        \textit{General Fixture} & Extract method & 4 \\
        \textit{Lack Of Cohesion Of Test Methods} & Extract class & 3 \\
        \textit{Indirect Testing} & Extract method & 3 \\
        \textit{Unknown Test} & Add assertion & 2 \\
        \textit{Conditional Test Logic} & Extract method & 2 \\
        \textit{Test Run War} & Make resource unique & 2 \\
        \bottomrule
    \end{tabular}
    \end{scriptsize}
\end{table}

\rqtakeaway{RQ3 shows that refactoring evidence is concentrated on a limited subset of test smells and ecosystems, with \textit{Extract Method} emerging as the dominant strategy across the literature. Detailed refactoring-by-language evidence is provided in~\ref{ap:test-smells-refactorings}.}

\subsection{RQ4: What datasets of test smells are available for each programming language and framework?}

To answer this question, we collected the test smell datasets mentioned or used in the referenced works. We manually examined these datasets to verify their availability. After this process, we identified a total of 16 test smell datasets. These datasets cover the following languages/frameworks: C++/GoogleTest, C\#, C\#/xUnit, Java, Java/JUnit, Python/Unittest, and Robot Framework. In total, the datasets encompass 73 different test smells across these seven languages/frameworks. The list of test smells in each dataset is available in the replication package.
\ref{ap:rq4-datasets} (Table~\ref{tab:test-smells-dataset}) provides the full dataset catalog and URLs.

\begin{lstlisting}[caption=Refactored Assertion Roulette Java example. Source: Adapted from \citet{Peruma2020}, label={lst:refactored-assertion-roulette-java-example-extract-method}]
@Test
public void testRepositoryContainsExpectedGitObject() throws Exception {
    Clone cloneOp = new Clone(
        false,
        integrationGitServerURIFor("small-repo.early.git"),
        helper().newFolder()
    );
    Repository repo = executeAndWaitFor(cloneOp);
    assertThat(
        repo,
        hasGitObject("ba1f63e4430bff267d112b1e8afc1d6294db0ccc")
    );
}

@Test
public void testReadmeFileExists() throws Exception {
    Clone cloneOp = new Clone(
        false,
        integrationGitServerURIFor("small-repo.early.git"),
        helper().newFolder()
    );
    Repository repo = executeAndWaitFor(cloneOp);
    File readmeFile = new File(repo.getWorkTree(), "README");
    assertThat(readmeFile, exists());
}

@Test
public void testReadmeFileHasCorrectLength() throws Exception {
    Clone cloneOp = new Clone(
        false,
        integrationGitServerURIFor("small-repo.early.git"),
        helper().newFolder()
    );
    Repository repo = executeAndWaitFor(cloneOp);
    File readmeFile = new File(repo.getWorkTree(), "README");
    assertThat(
        readmeFile.length(),
        equalTo(12L)
    );
}
\end{lstlisting}

\rqtakeaway{RQ4 reveals that publicly available datasets exist for only a limited subset of language and framework contexts, with a particularly strong concentration on Java and Java/JUnit.}

\subsection{RQ5: Which test smells are most critical across different programming languages and frameworks?}


To answer this question, we collected evidence directly from the papers regarding the criticality of different test smells. Our goal was not to produce a single authoritative ranking of test smell criticality, but rather to synthesize how the literature evaluates and reports it. Therefore, our assessment was grounded in the authors' own discussions, observations, or empirical metrics, rather than in our subjective judgment. \ref{ap:rq5-criticality-studies} lists all studies from which we extracted criticality information (Table~\ref{tab:rq5-criticality-by-study}).

After gathering criticality evidence from multiple studies, we standardized these assessments by organizing them into a five-level scale ranging from \textit{very low} to \textit{very high} criticality. This normalization enabled us to compare how primary studies report criticality and to identify points of agreement and disagreement in the literature.

\subsubsection{Analysis Criteria}

For each paper that discussed the severity or impact of test smells, we extracted the criticality assessment as reported by the authors (verbatim or near-verbatim), together with the paper ID and the location in the paper (e.g., section, table, or figure). We did not impose our own judgment; instead, we recorded the authors' original labels, scales, and rationales.

Primary studies assessed criticality from different perspectives. We classified each assessment into one of the following categories: (1) \textit{Developer perception in general}: how developers perceive the severity of the smell; (2) \textit{Impact on maintainability}: the effect of the smell on the maintainability of tests or code; (3) \textit{Impact on usefulness in development tasks}: the relevance of the smell to development-related tasks; (4) \textit{Not clear}: cases in which the paper did not explicitly specify the perspective used to assess criticality; and (5) \textit{Energy consumption}: one study (S107) classified test smells according to their impact on energy usage.

To enable comparison across studies that used different terminology and rating scales (e.g., ``Irrelevant,'' ``Not harmful,'' ``Highly harmful (5/5),'' ``Very High (5/5)''), we mapped these heterogeneous labels onto a unified five-level scale: \textit{very low}, \textit{low}, \textit{moderate}, \textit{high}, and \textit{very high}. Table~\ref{tab:criticality-mapping} presents the mapping rules applied in this process.

\begin{table}[h]
    \scriptsize
    \centering
    \caption{Mapping from primary-study severity labels to the five-level synthesis scale (used only for cross-study comparison).}
    \label{tab:criticality-mapping}
    \begin{tabular}{ll}
        \toprule
        \textbf{Standardized Scale} & \textbf{Examples of Labels Used in Primary Studies} \\
        \midrule
        Very low & Irrelevant (1/5), Not harmful, Very low severity, Less critical \\
        Low & Few severe (2/5), Low (1/3--2/5), Less severe than others \\
        Moderate & Medium (3/5), Low--medium, Moderate (2/3) \\
        High & High (4/5), More severe than others, High impact, One of the most harmful \\
        Very high & Highly harmful (5/5), Very high (5/5), Very severe (5/5), More critical \\
        \bottomrule
    \end{tabular}
\end{table}

We excluded the following cases from the main analysis: (i) entries in which the subject of criticality was classified as \textit{Not clear}; (ii) the \textit{energy-consumption} perspective, as it was reported by only a single study; and (iii) duplicate or overlapping assessments derived from the same underlying study (e.g., when a paper reported the criticality of a smell by citing another paper that we had already analyzed), as documented in our extraction notes.

In addition, test smells for which criticality evidence was reported by only one paper were not emphasized in the qualitative analysis due to the limited amount of supporting evidence. For this reason, smells classified under the perspective of \textit{impact on usefulness in development tasks} were not analyzed in the main discussion, since their criticality was reported in only one study.

\subsubsection{Analysis Results}

The test smells for which we collected criticality evidence from multiple papers include \textit{Assertion Roulette}, \textit{Conditional Test Logic}, \textit{Duplicate Assert}, \textit{Eager Test}, \textit{Ignored Test}, \textit{Magic Number}, \textit{Mystery Guest}, \textit{Obscure Test}, \textit{Resource Optimism}, and \textit{Sleepy Test}, among others.

A central finding is that primary studies employ different scales and evaluation criteria, and often disagree substantially on the severity of the same test smell. We illustrate this with three examples drawn directly from our data extraction (paper IDs refer to the list in~\ref{ap:selected-papers}).

\textit{Eager Test} exhibits particularly strong inconsistency. From a developer-perception perspective, Study S2 classifies it as ``Highly harmful (5/5),'' whereas Study S98 labels it ``Not harmful.'' From a maintainability perspective, Study S60 describes it as ``One of the most harmful,'' while Study S67 rates it as ``Very low (1/5)'' and Study S81 reports ``Very low severity.'' Thus, the same smell is described at both extremes of severity depending on the study.

\textit{Assertion Roulette} provides another illustrative example. Studies classify it as ``Irrelevant (1/5)'' (S2), ``Few severe (2/5)'' (S7), ``Less severe than others'' (S24), ``Medium (3/5)'' (S67), ``Less critical'' (S16), and ``Very low severity'' (S81). These assessments range from irrelevant to medium severity, indicating a lack of consensus in the literature.

\textit{Obscure Test}, when evaluated from a maintainability perspective, is classified as ``Low (2/5)'' in Study S34 but ``High (4/5)'' in Study S67, representing a direct contradiction within the same evaluation perspective.

Despite these disagreements, some patterns can still be observed. From a maintainability perspective, \textit{Sleepy Test}, \textit{Ignored Test}, \textit{Resource Optimism}, and \textit{Mystery Guest} are frequently reported as having high or very high criticality (e.g., S24, S34, S67). In contrast, \textit{Assertion Roulette} and \textit{Eager Test}, although among the most frequently studied smells, exhibit the widest range of ratings (from very low to high), indicating a lack of consensus in the literature. Meanwhile, \textit{Magic Number} is consistently reported as having low or very low criticality from a developer-perception perspective (S2, S98).

\begin{figure*}[!h]
    \centering
    \includegraphics[width=\textwidth]{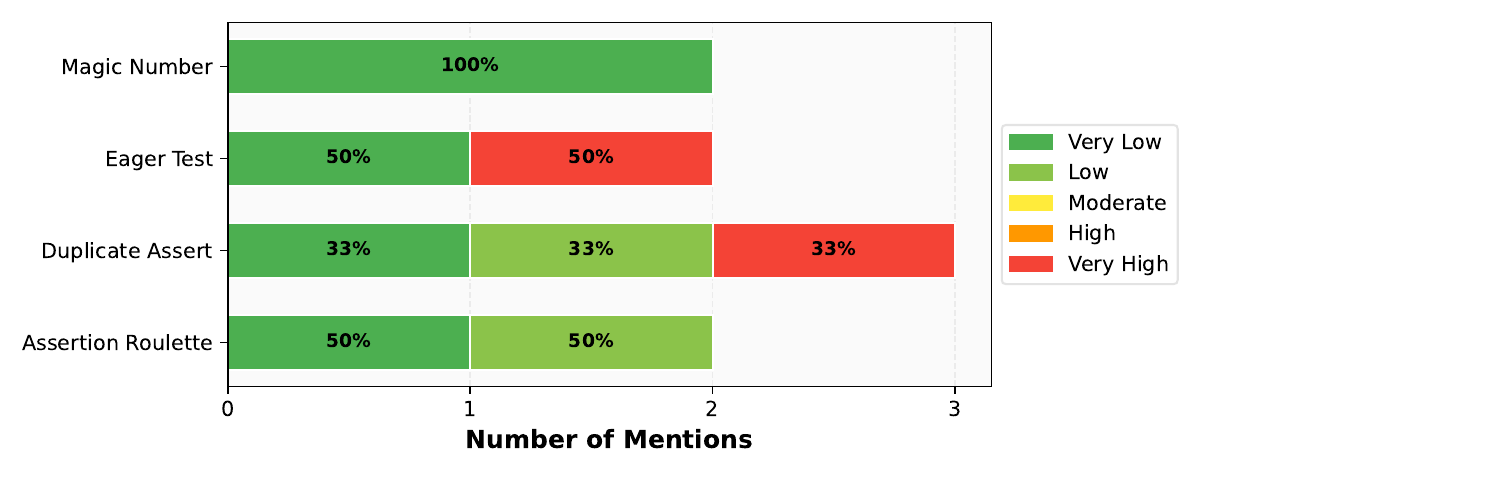}
    \caption{Test Smells Criticality Concerning to Developer's Perception in General.}
    \label{fig:criticality-developers-perception}
\end{figure*}

\begin{figure*}[!h]
    \centering
    \includegraphics[width=\textwidth]{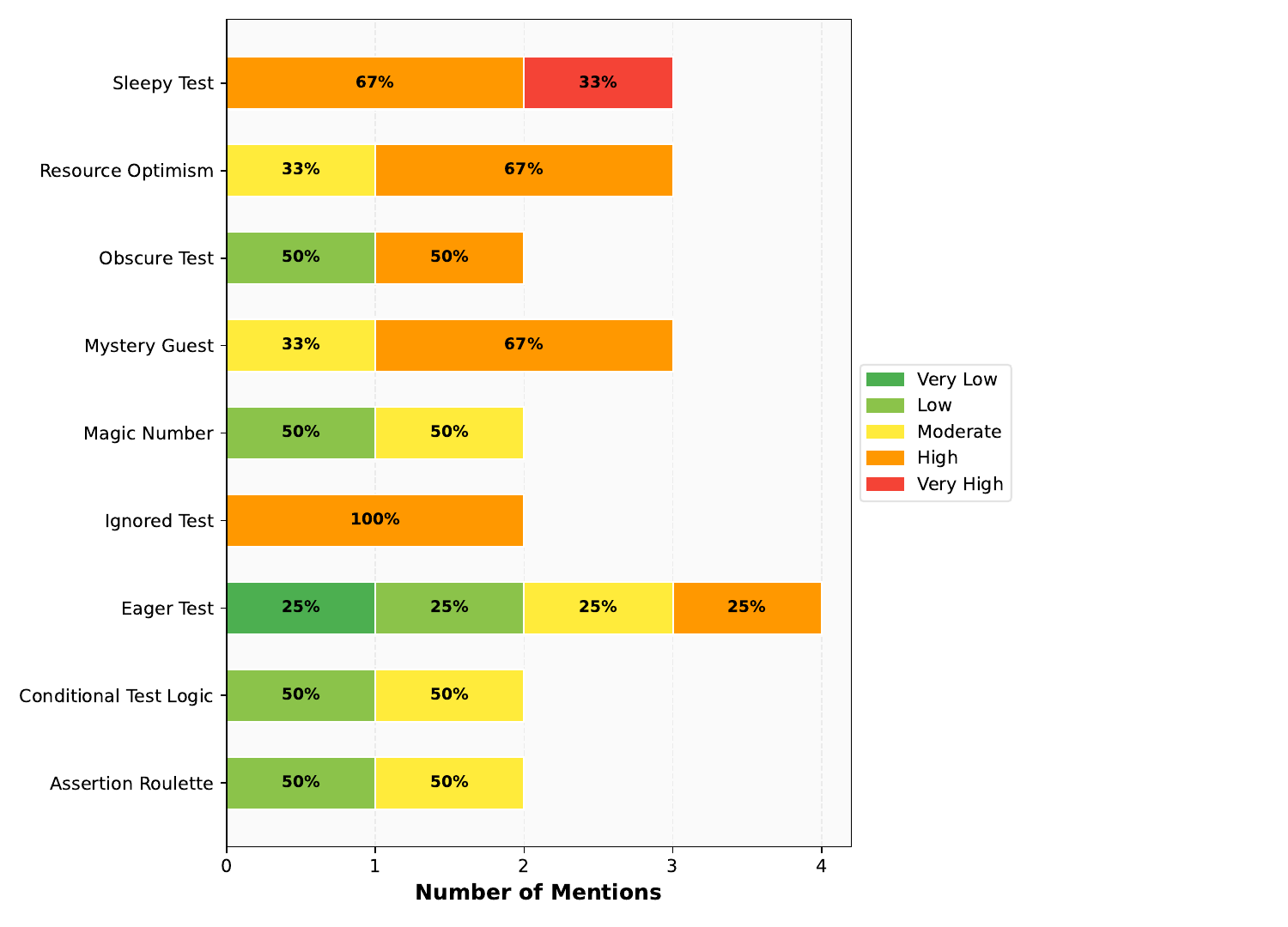}
    \caption{Test Smells Criticality Concerning to Maintainability.}
    \label{fig:criticality-maintainability}
\end{figure*}

An illustrative synthesis of how the extracted labels were mapped to the five-level scale for a subset of smells (considering developer perception and maintainability) is presented in Figures~\ref{fig:criticality-developers-perception} and~\ref{fig:criticality-maintainability}. These figures highlight how different studies rate the criticality of the same test smells.

\rqtakeaway{RQ5 indicates that criticality reporting is heterogeneous and often contradictory across studies. Nevertheless, certain smells (such as \textit{Sleepy Test}, \textit{Ignored Test}, \textit{Resource Optimism}, and \textit{Mystery Guest}) consistently emerge as high-risk from a maintainability perspective.}

\subsection{RQ6: What test smell detection or refactoring tools have been proposed for each programming language and framework?}

To answer this question, we collected tools related to test smells mentioned in the literature, including extensions to existing tools. In total, we identified 36 tools for C\#, Java, JavaScript, TypeScript, Python, Scala, and TTCN-3. Of these, 22 tools (61\%) are currently available, while 14 (39\%) are unavailable (e.g., Better Code Hub Extension, Evosuite Extension, TASTE, TestComet \& Teslo, TSGame, TestHound, DFD, DFI, STEEL, and Pysta). Java has the most significant number of available tools (15), followed by Python (4). We also found tools designed for specialized testing languages, such as Ukwikora and sonar-ikora-plugin for Robot Framework, and TRex for TTCN-3. However, only one tool, STEEL, is available for JavaScript, highlighting a notable gap for that language. The full tool inventory is listed in~\ref{ap:rq6-tools} (Table~\ref{tab:detecting-refactoring-tools}).

Refactoring capabilities within the ecosystem remain limited: only five tools (RAIDE, TESTLER, DARTS, TRex, and UTRefactor) support both detection and automatic refactoring, accounting for just 14\% of the available solutions. This fact highlights a broader gap in the field, namely the need for tools that not only identify test smells but also provide automated refactoring support. A particularly noteworthy case is UTRefactor, introduced in one of the most recent studies (2025), which leverages a large language model (LLM) to refactor test smells. This approach demonstrates considerable potential, yet it remains largely underexplored.

\rqtakeaway{RQ6 reveals a tool landscape that is strongly centered on Java and focused primarily on detection rather than automated remediation. In contrast, multi-language support and automated refactoring remain relatively scarce, highlighting a significant opportunity for future research.}

\section{Discussion}
\label{sec:discussion}

This systematic mapping study provides several insights into the current state of test smell research and outlines potential directions for future investigations. Based on the analysis of 117 papers published between 2006 and 2025, we identified noteworthy patterns that reveal limitations in the current body of work while also highlighting promising avenues for further exploration.

\textbf{Takeaway 1: The Java Monoculture Problem.}  
Our results clearly indicate the existence of a \emph{Java monoculture} within test smell research, as 85 out of the 117 papers focus exclusively on Java. This fact presents both opportunities and challenges. On one hand, the emphasis on Java has deepened our understanding of how test smells manifest in this language, which may also inform insights for similar object-oriented languages. On the other hand, this narrow focus may not accurately reflect real-world software development practices, where developers commonly employ a variety of languages, each with its own unique characteristics.

A temporal analysis reveals the same trend. Although the use of other programming languages in test smell research has grown over the years, Java’s dominance has not diminished. In fact, its use has steadily increased. Some languages, such as Python, have recently gained more attention (e.g., eight papers published in 2024 alone). However, other widely used industrial languages (such as JavaScript, TypeScript, Go, and Rust) remain underrepresented. This gap between research focus and industry practice may limit the generalizability of current findings. Therefore, further studies should prioritize underexplored languages to provide a more comprehensive and representative understanding of test smells.

\textbf{Takeaway 2: Language Characteristics Influence Test Smell Patterns.}  
Analyzing test smell categories across languages with different characteristics reveals patterns suggesting that language design decisions may influence the types of smells that occur most frequently. For instance, in dynamically typed languages such as Python, TypeScript, and JavaScript, test smells from the \emph{test execution–behavior} category appear more often (11.9\% for Python, 22.2\% for TypeScript, and 14.3\% for JavaScript) compared to statically typed languages like Java, where the occurrence is lower (6.3\%). 

Another striking example is TTCN-3, a language specifically designed for writing tests, in which most test smell occurrences are \emph{code-related} (75.7\%). Similarly, Python’s permissive syntax leads to higher proportions of test smells categorized as \emph{code-related} (26.9\%) and \emph{issues in test steps} (28.4\%).

These observations reinforce the previous takeaway: expanding research to encompass more diverse programming languages is crucial. Current tools and techniques (mainly developed for Java) may not accurately capture the nuances or address the unique challenges presented by other languages.

\textbf{Takeaway 3: The Criticality Paradox.}  

Another particularly concerning finding from this systematic mapping study is the disconnect between the test smells most frequently studied in research and those perceived as most critical in practice. \textit{Assertion Roulette} stands out as one of the most extensively investigated test smells. However, it is consistently reported across studies as having low to moderate criticality, both in terms of its impact on maintainability and developers’ perceived severity. Conversely, \textit{Sleepy Test}, frequently reported in the literature as highly critical for maintainability, has received far less research attention.

This paradox suggests that future studies should focus more on practically impactful problems rather than continuing to emphasize well-studied but less severe smells. One possible explanation for this imbalance is that research has tended to prioritize test smells that are easier to detect, such as \textit{Assertion Roulette}, thereby overlooking more complex yet more critical ones like \textit{Resource Optimism}. Redirecting research efforts toward these underexplored, high-impact test smells would help ensure that future work aligns more closely with real-world testing challenges.

\textbf{Takeaway 4: Inconsistent Criticality Classifications.}  
A further issue related to criticality is the substantial disagreement among studies regarding how harmful specific test smells are, even when assessing the same aspect (such as maintainability). For instance, \textit{Eager Test} has been described as both “highly harmful” and “not harmful” in different studies, whereas \textit{Assertion Roulette} has been rated anywhere from “irrelevant” to “medium severity.”

These inconsistencies underscore the need for more rigorous and standardized research on the criticality of test smells. Future studies should employ larger sample sizes, adopt consistent evaluation frameworks, and use quantitative methodologies to improve the reliability and reproducibility of criticality assessments. Such methodological rigor is essential for building a more cohesive understanding of how test smells truly affect software quality.

\textbf{Takeaway 5: The Detection–Refactoring Imbalance.}  
Regarding tools related to test smells, our study reveals a significant imbalance between detection and refactoring solutions. While detection tools are abundant (31 identified), refactoring tools are scarce (only five identified). This disparity is concerning, as it shows that developers receive substantial support for detecting test smells but limited assistance for effectively addressing them. Bridging this gap enables developers to complement detection with actionable, automated, and reliable refactoring solutions.

\textbf{Takeaway 6: Underutilization of AI-Based Solutions.}  
Artificial intelligence (AI) has become an increasingly popular research topic in software engineering, yet its application within the context of test smells remains highly limited. Among all selected studies, only one (UTRefactor~\citep{Gao2025}) has explored the use of Large Language Models (LLMs) to automate test smell refactoring. This work demonstrates the promising potential of AI-driven approaches in this domain. However, AI remains scarcely adopted when compared to more traditional techniques discussed in Section~\ref{sec:background}. Expanding the use of AI for both detection and refactoring could open new opportunities for more intelligent and adaptive test smell handling.

\textbf{Takeaway 7: Limited Multi-Language Support.}  
Our results also highlight the need for tools that support multiple programming languages. Currently, only a few recent tools attempt to achieve this. For example, SniffML~\citep{Lopes2024} supports C++, C\#, and Java, representing a significant step toward broader language coverage. Language-agnostic solutions, such as SniffML, offer a promising approach to promoting inclusiveness and generalizability across diverse programming ecosystems in test smell research. 

\section{Threats To Validity}
\label{sec:threats-to-validity}

In this section, we describe the threats to validity according to the classification proposed by~\citet{wholin2000}.

\textbf{Internal validity} threats refer to the possibility of attributing the cause of an effect to a factor when actually a third, unconsidered factor is the actual cause. In the context of our SMS, this could occur if we incorrectly attribute patterns in test smell research to programming languages when other factors (such as research community focus, tool availability, or publication trends) might be the actual cause. To mitigate this threat, we carefully documented our search strategy and inclusion criteria, and multiple authors reviewed the study selection process to ensure consistent application of criteria. Inter-rater reliability statistics, such as Cohen's $\kappa$, were not computed for the study-selection phase. Instead, we followed systematic mapping guidelines for software engineering~\citep{Petersen2008, Petersen2015}, which emphasize clearly defined criteria, multi-reviewer screening, and structured resolution of disagreements, without explicitly requiring inter-rater reliability reporting. This approach is consistent with prior test-smell mapping studies~\citep{Garousi2015, Garousi2018, Aljedaani2021}.

\textbf{Construct validity} threats concern whether our measurements adequately represent the constructs we intend to measure. In our SMS, this relates to whether our search strategy and classification criteria accurately capture the concepts of "test smells," "programming languages," and "test frameworks" as intended. A potential threat is that different papers may define test smells differently, leading to inconsistent classification. To mitigate this, we conducted a pilot test with well-known reference works to validate our search string, and we established clear operational definitions for all key constructs. Additionally, we used consensus among multiple authors to resolve classification ambiguities. A related construct validity threat concerns the synthesis of refactoring techniques (RQ3). Some primary studies describe refactoring strategies through code examples or informal discussions without explicitly naming the underlying techniques. To avoid overinterpreting such implicit evidence, we recorded refactorings only when a technique was explicitly identified or clearly stated as addressing a specific test smell (Section~\ref{sec:research-methodology}). While this conservative extraction rule may result in the underrepresentation of refactorings described only implicitly, it reduces the risk of attributing refactorings that were not intended or claimed by the original authors.

\textbf{External validity} concerns the extent to which the study results can be generalized beyond the specific context of our research. In our SMS, this relates to whether our findings about test smell research patterns can be generalized to the broader software engineering community or to different time periods. A potential threat is that our search was limited to specific digital libraries and may not capture all relevant research. Additionally, the temporal scope of our study (ending in August 2025) may not reflect future trends. To mitigate this, we searched multiple major digital libraries (IEEE Xplore, ACM Digital Library, ScienceDirect, and SpringerLink), tested our search string iteratively across all databases, and used consensus among all authors to finalize the search strategy. We also documented our search process transparently to enable replication and extension.

\textbf{Conclusion validity} concerns whether the statistical tests and evidence adequately support the conclusions reached in the study. In our SMS, this relates to whether our analysis methods and data presentation support the claims we make about patterns in test smell research. A potential threat is that our descriptive analysis might not provide sufficient evidence for the conclusions drawn. To mitigate this, we used systematic data extraction procedures, provided detailed tables and figures to support our findings, and ensured that the extracted data directly support all conclusions. We also conducted a peer review of the analysis process to ensure consistency and accuracy.

\section{Related Work}
\label{sec:related-work}

In this section, we discuss relevant literature reviews on test smells, which were identified by consulting test smell experts.


\citet{Garousi2018} conducted a multivocal literature mapping on test smells, examining 166 sources from scientific and grey literature. The study provides practitioners with a detailed classification of test smells with recommendations for test automation, along with guidelines, techniques, and tools for their prevention, detection, and correction. Later, \citet{Garousi2019} conducted a similar review, analyzing 46 scientific and 120 grey literature sources to bridge the gap between academia and industry. Unlike these studies, which provide broad overviews of test smell knowledge from both scientific and grey literature, our work focuses on synthesizing the programming language and testing framework dimensions. More specifically, we analyze how test smells, refactorings, datasets, criticality evidence, and tools are distributed across programming ecosystems, and we distinguish between widely recurring smells and those that appear to be language-specific.

\citet{Tran2021} aimed to develop a comprehensive model to assess the quality of test cases and test suites. They performed a literature review of 49 secondary studies to achieve this goal. As a result, the paper presents a model to support the assessment and improvement of test artifact quality. The authors examined test smells and other quality attributes, whereas our work focuses solely on test smells. In our study, we did not propose any model for assessing test cases. However, our results can guide future studies aiming to develop such models, especially those focused on addressing test smells in multiple languages, since our work examined test smells from a programming language perspective.


\citet{Aljedaani2021} aimed to provide practitioners and researchers with a catalog of test smell detection tools. To achieve this goal, the authors conducted a literature review and identified 22 test smell detection tools. They further analyzed the test smells detected by each tool, the supported programming languages and frameworks, the detection strategies employed, and the extent to which each tool has been adopted. Their unit of analysis is therefore the tool itself and its characteristics. In contrast, our unit of analysis is the broader body of primary studies on test smells across programming languages and testing frameworks. Tools constitute only one of the six research-question dimensions in our systematic mapping study, alongside prevalence, universal versus language-specific smells, refactorings, datasets, and criticality.


The study by~\citet{Rwemalika2022} aims to address gaps in the research on test smells in System User Interactive Tests (SUITs). The authors conducted an exploratory analysis and performed a multivocal literature review, resulting in a catalog of 35 SUIT-specific test smells. They also analyzed the prevalence of these smells in both industrial and open-source projects. Unlike that domain-specific synthesis, our study does not restrict its scope to SUITs. Instead, it synthesizes the broader body of test smell literature through the lens of programming languages and testing frameworks. However, some of the languages and frameworks identified in our study suggest that certain SUIT-specific smells may also be captured. For instance, Selenium WebDriver (one of the frameworks identified in our work) is commonly used for developing SUITs.

\section{Conclusions}
\label{sec:conclusions}


In this work, we conducted a systematic mapping study to examine test smells through the lens of programming languages and test frameworks. We analyzed 117 papers and extracted data to address six research questions covering prevalence, universal versus language-specific smells, refactorings, datasets, criticality, and tools. Rather than proposing a new canonical catalog of test smells, our main contribution is a cross-language synthesis of the evidence reported in the literature. This synthesis makes explicit how test smell knowledge is distributed across programming ecosystems and where current research remains concentrated.

This study provides substantial value to both the research community and industry practitioners. For researchers, the results highlight where new studies are most needed: the literature remains concentrated on Java, while several other languages and frameworks are still underexplored, and the smells that receive the most attention are not always those perceived as most critical in practice. For tool builders, the consolidated mapping of 94 refactorings, 16 datasets, and 36 tools offers a concrete foundation for selecting benchmark data, identifying unsupported smells, and addressing gaps such as the imbalance between detection and refactoring support, as well as the limited availability of multi-language solutions. For practitioners, the study provides practical guidance for prioritizing improvement efforts. They can leverage the criticality synthesis and the language-specific mappings of refactorings and tools to focus on smells with stronger evidence of impact, as well as on remediation options that better align with their own ecosystems. Together, these contributions position this study as a cross-language evidence map for test smell research, complementing existing catalogs and tool-focused mapping studies rather than duplicating them.

For future work, we plan to propose a language-independent approach to detecting test smells and develop a tool that supports multiple languages and frameworks. This tool will be compared against existing solutions to address a need identified in this SMS. Additionally, we aim to create a new tool that facilitates the refactoring of test smells across various languages, as our findings indicate that most existing tools focus solely on detection. Another avenue for future research involves investigating test smells identified exclusively in specific languages (particularly those in Java) to determine whether they can also appear in other languages and frameworks. For extension or replication purposes, we make the replication package~\footnote{\url{https://zenodo.org/records/17093675}} of this work publicly available online.

\section*{Acknowledgements}

The authors gratefully acknowledge the financial support provided for this research. Márcio Ribeiro was partially supported by INES.IA (National Institute of Science and Technology for Software Engineering Based on and for Artificial Intelligence) (www.ines.org.br), under CNPq grant 408817/2024-0, as well as by CNPq grants 312195/ 2021-4, 404825/2023-0, and 443393/2023-0. Ivan Machado acknowledges partial support from the CAPES – Finance Code 001; Fundação de Amparo à Pesquisa do Estado da Bahia (FAPESB), grant PIE0002/2022; and CNPq, grants 315840/2023-4 and 403361/2023-0. Carla Bezerra is supported by 403304/2025-3.

\section*{Declaration of generative AI and AI-assisted technologies in the writing process}

During the preparation of this work the author(s) used ChatGPT in order to correct any syntax and grammatical errors in the text. After using this tool/service, the author(s) reviewed and edited the content as needed and take(s) full responsibility for the content of the publication.

\bibliographystyle{elsarticle-harv.bst} 
\bibliography{references.bib}

\FloatBarrier
\appendix

\section{Selected Papers\label{ap:selected-papers}}

\begin{scriptsize}
\begin{longtable}[c]{p{1cm}p{12cm}}
    \caption{Selected Papers} \label{tab:selected-papers} \\
    \toprule
    \textbf{ID} & \textbf{Title} \\
    \midrule
    \endfirsthead
    \multicolumn{2}{c}%
    {{\bfseries Continued from previous page}} \\
    \toprule
    \textbf{ID} & \textbf{Title} \\
    \midrule
    \endhead
    \midrule
    \multicolumn{2}{r}{{Continued on next page}} \\
    \endfoot
    \bottomrule
    \endlastfoot
    S1 & The Lost World: Characterizing and Detecting Undiscovered Test Smells~\citep{Yang2024} \\
    S2 & Analyzing Test Smells Refactoring from a Developers Perspective~\citep{Damasceno2023} \\
    S3 & TEMPY: Test Smell Detector for Python~\citep{Fernandes2022} \\
    S4 & Investigating Test Smells in JavaScript Test Code~\citep{Jorge2021} \\
    S5 & Refactoring Test Smells: A Perspective from Open-Source Developers~\citep{Soares2020} \\
    S6 & On the use of test smells for prediction of flaky tests~\citep{Camara2021} \\
    S7 & Understanding practitioners' strategies to handle test smells: a multi-method study~\citep{Santana2021} \\
    S8 & Handling Test Smells in Python: Results from a Mixed-Method Study~\citep{Fernandes2021} \\
    S9 & Investigating Developers' Contributions to Test Smell Survivability: A Study of Open-Source Projects~\citep{Campos2023} \\
    S10 & On the distribution of test smells in open source Android applications: an exploratory study~\citep{Peruma2019} \\
    S11 & On the influence of Test Smells on Test Coverage~\citep{Virginio2019} \\
    S12 & Do the Test Smells Assertion Roulette and Eager Test Impact Students' Troubleshooting and Debugging Capabilities?~\citep{Aljedaani2023} \\
    S13 & SoCRATES: Scala radar for test smells~\citep{DeBleser2019} \\
    S14 & PyTest-Smell: a smell detection tool for Python unit tests~\citep{Bodea2022} \\
    S15 & How Students Unit Test: Perceptions, Practices, and Pitfalls~\citep{Bai2021} \\
    S16 & Characterizing high-quality test methods: a first empirical study~\citep{Veloso2022} \\
    S17 & Identifying Test Refactoring Candidates with Assertion Fingerprints~\citep{Fang2015} \\
    S18 & LCCSS: A Similarity Metric for Identifying Similar Test Code~\citep{Silva2020} \\
    S19 & Fine-grained test minimization~\citep{Vahabzadeh2018} \\
    S20 & An empirical investigation on the readability of manual and generated test cases~\citep{Grano2018} \\
    S21 & Web test dependency detection~\citep{Biagiola2019} \\
    S22 & Modeling readability to improve unit tests~\citep{Daka2015} \\
    S23 & An empirical evaluation of RAIDE: A semi-automated approach for test smells detection and refactoring~\citep{Santana2024} \\
    S24 & On the diffusion of test smells and their relationship with test code quality of Java projects~\citep{Martins2024.1} \\
    S25 & Automatic Refactoring Method to Remove Eager Test Smell~\citep{Pizzini2023} \\
    S26 & Refactoring Test Smells With JUnit 5: Why Should Developers Keep Up-to-Date?~\citep{Soares2023} \\
    S27 & Refactoring Assertion Roulette and Duplicate Assert test smells: a controlled experiment~\citep{Santana2022} \\
    S28 & Sentinel: A process for automatic removing of Test Smells~\citep{Pizzini2023.2} \\
    S29 & Hearing the voice of experts: Unveiling Stack Exchange communities' knowledge of test smells~\citep{Martins2023} \\
    S30 & Do you see any problem? On the Developers Perceptions in Test Smells Detection~\citep{Lima2023} \\
    S31 & PYNOSE: A test smell detector for python~\citep{Wang2021} \\
    S32 & Machine Learning-Based Test Smell Detection~\citep{Pontillo2024} \\
    S33 & JTDog: a Gradle Plugin for Detecting Dynamic Test Smells~\citep{Taniguchi2021} \\
    S34 & Investigating developer perception on test smells using better code hub - Work in progress -~\citep{Schvarcbacher2019} \\
    S35 & How and why developers implement OS-specific tests~\citep{Job2024} \\
    S36 & An Empirical Study on the Evolution of Test Smell~\citep{Kim2020} \\
    S37 & RAIDE: A tool for Assertion Roulette and Duplicate Assert identification and refactoring~\citep{Santana2020} \\
    S38 & An Empirical Study of Using Large Language Models for Unit Test Generation~\citep{Schäfer2024} \\
    S39 & An Exploratory Study on the Refactoring of Unit Test Files in Android Applications~\citep{Peruma2020.2} \\
    S40 & Automatic Generation of Smell-free Unit Tests~\citep{Afonso2023} \\
    S41 & A manual categorization of new quality issues on automatically-generated tests~\citep{Galindo-Gutierrez2023} \\
    S42 & On the detection of test smells: A metrics-based approach for general fixture and eager test~\citep{VanRompaey2007} \\
    S43 & How disabled tests manifest in test maintainability challenges?~\citep{Kim2021} \\
    S44 & To What Extent Can Code Quality be Improved by Eliminating Test Smells?~\citep{Wu2022} \\
    S45 & Just-In-Time Test Smell Detection and Refactoring: The DARTS Project~\citep{Lambiase2020} \\
    S46 & A Trend Analysis of Test Smells in Python Test Code Over Commit History~\citep{Fushihara2023} \\
    S47 & A preliminary evaluation on the relationship among architectural and test smells~\citep{DeStefano2022} \\
    S48 & Assessing Diffusion and Perception of Test Smells in Scala Projects~\citep{DeBleser2019.2} \\
    S49 & On the Diffusion of Test Smells in Automatically Generated Test Code: An Empirical Study~\citep{Palomba2016} \\
    S50 & Automatic Test Smell Detection Using Information Retrieval Techniques~\citep{Palomba2018} \\
    S51 & Revisiting Test Smells in Automatically Generated Tests: Limitations, Pitfalls, and Opportunities~\citep{Panichella2020} \\
    S52 & An Exploratory Study of the Relationship Between Software Test Smells and Fault-Proneness~\citep{Qusef2019} \\
    S53 & An Analysis of Complex Industrial Test Code Using Clone Analysis~\citep{Hasanain2018} \\
    S54 & Automated Detection of Test Fixture Strategies and Smells~\citep{Greiler2013.1} \\
    S55 & Generated Tests in the Context of Maintenance Tasks: A Series of Empirical Studies~\citep{Herculano2022} \\
    S56 & Revisiting Test Impact Analysis in Continuous Testing From the Perspective of Code Dependencies~\citep{Peng2022} \\
    S57 & Why is Unit-testing in Computer Games Difficult?~\citep{Toll2012} \\
    S58 & Characterizing the Relative Significance of a Test Smell~\citep{VanRompaey2006} \\
    S59 & On the Distribution of ``Simple Stupid Bugs'' in Unit Test Files: An Exploratory Study~\citep{Peruma2021} \\
    S60 & Scented since the beginning: On the diffuseness of test smells in automatically generated test code~\citep{GRANO2019} \\
    S61 & Why do builds fail?—A conceptual replication study~\citep{BARRAK2021} \\
    S62 & A multi-label classification approach for detecting test smells over java projects~\citep{HADJKACEM2022} \\
    S63 & A comparative study of test code clones and production code clones~\citep{VANBLADEL2021} \\
    S64 & What do developers consider magic literals? A smalltalk perspective~\citep{ANQUETIL2022} \\
    S65 & The secret life of test smells - an empirical study on test smell evolution and maintenance~\citep{Kim2021.2} \\
    S66 & Use of Test Doubles in Android Testing: An In-Depth Investigation~\citep{Fazzini2022} \\
    S67 & Investigating Severity Thresholds for Test Smells~\citep{Spadini2020} \\
    S68 & Strategies for avoiding text fixture smells during software evolution~\citep{Greiler2013.2} \\
    S69 & Randoop-TSR: Random-based Test Generator with Test Suite Reduction~\citep{Liu2022} \\
    S70 & JNose: Java Test Smell Detector~\citep{Virginio2020} \\
    S71 & An empirical study of automatically-generated tests from the perspective of test smells~\citep{Virginio2020.2} \\
    S72 & Automatically identifying dead fields in test code by resolving method call and field dependency~\citep{Satter2017} \\
    S73 & An empirical investigation into the nature of test smells~\citep{Tufano2016} \\
    S74 & Smart prediction for refactorings in the software test code~\citep{Martins2021} \\
    S75 & A comprehensive catalog of refactoring strategies to handle test smells in Java-based systems~\citep{Martins2024.2} \\
    S76 & An empirical study on the influence of developers' experience on software test code quality~\citep{Campos2023.2} \\
    S77 & Detection of test smells with basic language analysis methods and its evaluation~\citep{Maier2023} \\
    S78 & Ukwikora: Continuous inspection for keyword-driven testing~\citep{Rwemalika2019} \\
    S79 & TsDetect: An open source test smells detection tool~\citep{Peruma2020} \\
    S80 & A static code search technique to identify dead fields by analyzing usage of setup fields and field dependency in test code~\citep{Satter2016} \\
    S81 & Test smells 20 years later: detectability, validity, and reliability~\citep{Panichella2022} \\
    S82 & Who Is Afraid of Test Smells? Assessing Technical Debt from Developer Actions~\citep{Chen2023} \\
    S83 & Smells in system user interactive tests~\citep{Rwemalika2022} \\
    S84 & Evaluating test quality of Python libraries for IoT applications at the network edge~\citep{Chen2024} \\
    S85 & Investigating developers' perception on software testability and its effects~\citep{Sharma2023} \\
    S86 & Are test smells really harmful? An empirical study~\citep{Bavota2015} \\
    S87 & Pragmatic Approach to Test Case Reuse - A Case Study in Android OS BiDiTests Library~\citep{Asaithambi2014} \\
    S88 & Towards Test Case Reuse: A Study of Redundancies in Android Platform Test Libraries~\citep{Asaithambi2013} \\
    S89 & An approach to quality engineering of TTCN-3 test specifications~\citep{Neukirchen2008} \\
    S90 & JExample: Exploiting Dependencies between Tests to Improve Defect Localization~\citep{Kuhn2008} \\
    S91 & Utilising Code Smells to Detect Quality Problems in TTCN-3 Test Suites~\citep{Neukirchen2007} \\
    S92 & Using Large Language Models to Generate JUnit Tests: An Empirical Study~\citep{Siddiq2024} \\
    S93 & Understanding Test Convention Consistency as a Dimension of Test Quality~\citep{Robillard2024} \\
    S94 & Beyond Accuracy: An Empirical Study on Unit Testing in Open-source Deep Learning Projects~\citep{Wang2024} \\
    S95 & Using GitHub Copilot for Test Generation in Python: An Empirical Study~\citep{ElHaji2024} \\
    S96 & From Fine-tuning to Output: An Empirical Investigation of Test Smells in Transformer-Based Test Code Generation~\citep{Aljohani2024} \\
    S97 & Test Smells Learning by a Gamification Approach~\citep{Fasolino2024} \\
    S98 & Insights from the Field: Exploring Students' Perspectives on Bad Unit Testing Practices~\citep{Peruma2024} \\
    S99 & XNose: A Test Smell Detector for C\#~\citep{Paul2024} \\
    S100 & What's in a Display Name? An Empirical Study on the Use of Display Names in Open-Source JUnit Tests~\citep{Qiao2024} \\
    S101 & Toward granular search-based automatic unit test case generation~\citep{Pecorelli2024} \\
    S102 & A Road to Find Them All: Towards an Agnostic Strategy for Test Smell Detection~\citep{Lopes2024} \\
    S103 & How Aware Are We of Test Smells in Quantum Software Systems? A Preliminary Empirical Evaluation~\citep{Virginio2024} \\
    S104 & Automated Unit Test Refactoring~\citep{Gao2025} \\
    S105 & Gamifying Testing in IntelliJ: A Replicability Study~\citep{Straubinger2025} \\
    S106 & Sojourner under Sabotage: A Serious Testing and Debugging Game~\citep{Straubinger2025.2} \\
    S107 & Test smell: A parasitic energy consumer in software testing~\citep{Misu2025} \\
    S108 & Predicting the Root Cause of Flaky Tests Based on Test Smells~\citep{Wang2025} \\
    S109 & Understanding the Prevalence of Test Smells in Open-source and Industrial Software: An Empirical Study on Python Projects~\citep{Hasan2024} \\
    S110 & Fault-Proneness of Python Programs Tested By Smelled Test Code~\citep{Fushihara2024} \\
    S111 & Private-Keep Out? Understanding How Developers Account for Code Visibility in Unit Testing~\citep{Roslan2024} \\
    S112 & Viscount: A Direct Method Call Coverage Tool for Java~\citep{Roslan2024.2} \\
    S113 & Reinforcement Learning from Automatic Feedback for High-Quality Unit Test Generation~\citep{Steenhoek2025} \\
    S114 & How Do Developers Structure Unit Test Cases? An Empirical Analysis of the AAA Pattern in Open Source Projects~\citep{Wei2025} \\
    S115 & A proposal and assessment of an improved heuristic for the Eager Test smell detection~\citep{Tran2025} \\
    S116 & An empirical investigation into the capabilities of anomaly detection approaches for test smell detection~\citep{Pontillo2025} \\
    S117 & Automated Clone Elimination in Python Tests~\citep{Kingston2025} \\
\end{longtable}
\end{scriptsize}

\section{RQ1 language and framework frequencies\label{ap:rq1-languages-frameworks}}

\begin{table}[H]
    \scriptsize
    \centering
    \caption{Languages and Frameworks}
    \label{tab:languages-frameworks}
    \begin{tabular}{lll}
        \toprule
        \textbf{Language} & \textbf{Framework} & \textbf{Number of Papers} \\
        \midrule
        C & - & 1 \\
        C++ & - & 1 \\
        C++ & GoogleTest & 1 \\
        C\# & - & 1 \\
        C\# & xUnit & 1 \\
        Java & JUnit & 62 \\
        Java & - & 27 \\
        Java & TestNG & 4 \\
        Java & Selenium WebDriver & 1 \\
        JavaScript & Chai & 1 \\
        JavaScript & Jest & 1 \\
        Kotlin & - & 1 \\
        Python & Unittest & 10 \\
        Python & PyTest & 8 \\
        Python & Nose & 2 \\
        Python & - & 1 \\
        Python & absl & 1 \\
        Python & GoogleTest & 1 \\
        Python & Numpy & 1 \\
        Python & Tensorflow & 1 \\
        Scala & ScalaTest & 3 \\
        Smalltalk & - & 1 \\
        TTCN-3 & - & 2 \\
        Typescript & Jest & 2 \\
        Typescript & Chai & 1 \\
        - & Robot Framework & 2 \\
        \bottomrule
    \end{tabular}
\end{table}

\section{RQ1 top cited test smells\label{ap:rq1-top-smells}}

\begin{table}[H]
    \scriptsize
    \centering
    \caption{Top 20 Most Cited Test Smells Across All Languages}
    \label{tab:top-20-test-smells}
    \begin{tabular}{p{3cm}cccccccccc}
        \toprule
        \textbf{Test Smell} & \textbf{C} & \textbf{C\#} & \textbf{C++} & \textbf{Java} & \textbf{Kotlin} & \textbf{Python} & \textbf{Scala} & \textbf{JS} & \textbf{TS} & \textbf{All} \\
        \midrule
        \textit{Assertion Roulette} & 0 & 2 & 1 & 49 & 0 & 7 & 2 & 1 & 1 & 63 \\
        \textit{Eager Test} & 0 & 1 & 0 & 51 & 0 & 2 & 2 & 1 & 1 & 59 \\
        \textit{Conditional Test Logic} & 0 & 3 & 1 & 31 & 0 & 10 & 0 & 2 & 1 & 48 \\
        \textit{Duplicate Assert} & 0 & 3 & 1 & 32 & 0 & 7 & 0 & 1 & 1 & 45 \\
        \textit{Exception Handling} & 0 & 2 & 1 & 27 & 0 & 10 & 0 & 1 & 1 & 42 \\
        \textit{Magic Number} & 0 & 2 & 2 & 28 & 0 & 7 & 0 & 1 & 1 & 41 \\
        \textit{Sensitive Equality} & 0 & 1 & 0 & 37 & 0 & 1 & 2 & 0 & 0 & 41 \\
        \textit{Mystery Guest} & 0 & 0 & 0 & 36 & 0 & 1 & 2 & 1 & 1 & 41 \\
        \textit{Unknown Test} & 0 & 3 & 1 & 26 & 0 & 7 & 0 & 1 & 1 & 39 \\
        \textit{General Fixture} & 0 & 0 & 0 & 32 & 1 & 4 & 2 & 0 & 0 & 39 \\
        \textit{Sleepy Test} & 0 & 1 & 0 & 23 & 0 & 8 & 0 & 1 & 1 & 34 \\
        \textit{Resource Optimism} & 0 & 0 & 0 & 31 & 0 & 1 & 0 & 1 & 1 & 34 \\
        \textit{Empty Test} & 0 & 2 & 1 & 22 & 0 & 7 & 0 & 0 & 0 & 32 \\
        \textit{Ignored Test} & 0 & 1 & 0 & 23 & 0 & 5 & 0 & 1 & 1 & 31 \\
        \textit{Redundant Assertion} & 0 & 1 & 0 & 21 & 0 & 6 & 0 & 1 & 1 & 30 \\
        \textit{Lazy Test} & 0 & 0 & 0 & 24 & 0 & 1 & 2 & 1 & 1 & 29 \\
        \textit{Redundant Print} & 0 & 1 & 0 & 12 & 0 & 7 & 0 & 2 & 1 & 23 \\
        \textit{Obscure Test} & 0 & 0 & 1 & 18 & 0 & 3 & 0 & 0 & 0 & 23 \\
        \textit{Constructor Initialization} & 0 & 1 & 0 & 16 & 0 & 5 & 0 & 0 & 0 & 22 \\
        \textit{Duplicated Test Code} & 1 & 0 & 0 & 11 & 0 & 0 & 0 & 0 & 0 & 13 \\
        \bottomrule
    \end{tabular}
\end{table}

\section{Test smells that appear in only one language or framework}
\label{ap:specific-test-smells}

\begin{scriptsize}
\begin{longtable}[c]{p{4cm}p{9cm}}
    \caption{Test Smells That Appear In Only One Language Or Framework} \label{tab:specific-test-smells} \\
    \toprule
    \textbf{Language/Framework} & \textbf{Test Smell} \\
    \midrule
    \endfirsthead
    \multicolumn{2}{c}%
    {{\bfseries Continued from previous page}} \\
    \toprule
    \textbf{Language/Framework} & \textbf{Test Smell} \\
    \midrule
    \endhead
    \midrule
    \multicolumn{2}{r}{{Continued on next page}} \\
    \endfoot
    \bottomrule
    \endlastfoot
    C++ & \textit{Fragile Test} \\
    Java & \textit{Assertion In Loop} \\
    Java & \textit{Dependent Assert} \\
    Java & \textit{Fire And Forget} \\
    Java & \textit{Multithreading} \\
    Java & \textit{Non-Deterministic Data Test} \\
    Java & \textit{Ordered Hypothesis} \\
    Java & \textit{Parameter Order Wrong} \\
    Java & \textit{Poor Test Organization} \\
    Java & \textit{Return In Test} \\
    Java & \textit{Time Acquisition} \\
    Java/JUnit & \textit{Ambiguous Display Name} \\
    Java/JUnit & \textit{Anal Probe} \\
    Java/JUnit & \textit{Annotation Repetition} \\
    Java/JUnit & \textit{Anonymous Test} \\
    Java/JUnit & \textit{Arrange\&Quit} \\
    Java/JUnit & \textit{Assert Pre-Condition} \\
    Java/JUnit & \textit{Asserting Constants} \\
    Java/JUnit & \textit{Asserting Object Initialization Multiple Times} \\
    Java/JUnit & \textit{Assertions With Not Related Parent Class Method} \\
    Java/JUnit & \textit{Chain Gang} \\
    Java/JUnit & \textit{Code Organization} \\
    Java/JUnit & \textit{Coupling Between Test Methods} \\
    Java/JUnit & \textit{Coupling With Production Code} \\
    Java/JUnit & \textit{Dependencies On Test Containers} \\
    Java/JUnit & \textit{Doppelgänger} \\
    Java/JUnit & \textit{Duplicate Display Names} \\
    Java/JUnit & \textit{Duplicated Setup} \\
    Java/JUnit & \textit{Exceptions Due To External Dependencies} \\
    Java/JUnit & \textit{Exceptions Due To Incomplete Setup} \\
    Java/JUnit & \textit{Exceptions Due To Null Arguments} \\
    Java/JUnit & \textit{For Testers Only} \\
    Java/JUnit & \textit{Friction} \\
    Java/JUnit & \textit{God Test Class} \\
    Java/JUnit & \textit{Happy Path Only} \\
    Java/JUnit & \textit{Hard To Co-Evolve} \\
    Java/JUnit & \textit{Hardcoded Dependency} \\
    Java/JUnit & \textit{Hidden Dependency} \\
    Java/JUnit & \textit{Inconsistent Display Name} \\
    Java/JUnit & \textit{Inconsistent Style} \\
    Java/JUnit & \textit{Interacting Test Suites} \\
    Java/JUnit & \textit{Irrelevant Information} \\
    Java/JUnit & \textit{Is Mockito Working Fine} \\
    Java/JUnit & \textit{Lack Comments} \\
    Java/JUnit & \textit{Likely Ineffective Object Comparison} \\
    Java/JUnit & \textit{Messy Tests} \\
    Java/JUnit & \textit{Missing Assert} \\
    Java/JUnit & \textit{Missing Display Name} \\
    Java/JUnit & \textit{Mock Call Verification} \\
    Java/JUnit & \textit{Mock Makes The Test Always Passes} \\
    Java/JUnit & \textit{Mock Naming} \\
    Java/JUnit & \textit{Multilingual Display Name} \\
    Java/JUnit & \textit{Multiple Aaa} \\
    Java/JUnit & \textit{Multiple Acts} \\
    Java/JUnit & \textit{Naming Convention Violation} \\
    Java/JUnit & \textit{No Assertions} \\
    Java/JUnit & \textit{Not Asserted Return Values} \\
    Java/JUnit & \textit{Not Asserted Side Effects} \\
    Java/JUnit & \textit{Obscure Assert} \\
    Java/JUnit & \textit{Obscure Inline Setup} \\
    Java/JUnit & \textit{Obsolete Tests} \\
    Java/JUnit & \textit{Overreferencing} \\
    Java/JUnit & \textit{Redundant Display Name} \\
    Java/JUnit & \textit{Redundant Not Null Assertion} \\
    Java/JUnit & \textit{Redundant Test} \\
    Java/JUnit & \textit{Refused Bequest} \\
    Java/JUnit & \textit{Replicating Disadvantages Of The Code} \\
    Java/JUnit & \textit{Resource Leakage} \\
    Java/JUnit & \textit{Resource Optimist} \\
    Java/JUnit & \textit{Second Class Citizens} \\
    Java/JUnit & \textit{State Leak} \\
    Java/JUnit & \textit{Suppressed Exception} \\
    Java/JUnit & \textit{Test Logic In Production Code} \\
    Java/JUnit & \textit{Test Run War} \\
    Java/JUnit & \textit{Testing Only Field Accessors} \\
    Java/JUnit & \textit{Testing The Same Exception Scenario} \\
    Java/JUnit & \textit{Testing The Same Void Method} \\
    Java/JUnit & \textit{The Butterfly} \\
    Java/JUnit & \textit{The Conjoined Twins} \\
    Java/JUnit & \textit{The Dead Tree} \\
    Java/JUnit & \textit{The Environmental Vandal} \\
    Java/JUnit & \textit{The Forty Foot Pole Test} \\
    Java/JUnit & \textit{The Giant} \\
    Java/JUnit & \textit{The Sleeper} \\
    Java/JUnit & \textit{The Test With No Name} \\
    Java/JUnit & \textit{Tooling Details In Test Case} \\
    Java/JUnit & \textit{Ugly Mirror} \\
    Java/JUnit & \textit{Unclear Display Name} \\
    Java/JUnit & \textit{Unecessary Test} \\
    Java/JUnit & \textit{Ungrammatical Display Name} \\
    Java/JUnit & \textit{Unrelated Assertions} \\
    Java/JUnit & \textit{Unused Inputs} \\
    Java/JUnit & \textit{Using More Than One Mock For A Test} \\
    Java/TestNG & \textit{Duplicate Test Runs Caused By Inheritance} \\
    Java/TestNG & \textit{Scattered Test Fixtures Caused By Inheritance} \\
    Java/TestNG & \textit{Using Test Case Inheritance To Test Source Code Polymorphism} \\
    Java/TestNG & \textit{Vague Header Setup Smell} \\
    Python/PyTest & \textit{Test Clone} \\
    Python/Unittest & \textit{Lack Of Cohesion Of Test Cases} \\
    Python/Unittest & \textit{Suboptimal Assert} \\
    Robot Framework & \textit{Army Of Clones} \\
    Robot Framework & \textit{Complicated Setup Scenarios} \\
    Robot Framework & \textit{Conditional Assertions} \\
    Robot Framework & \textit{Conspiracy Of Silence} \\
    Robot Framework & \textit{Data Creep} \\
    Robot Framework & \textit{Dependencies Between Tests} \\
    Robot Framework & \textit{Directly Executing Ui Scripts} \\
    Robot Framework & \textit{Duplicate Check} \\
    Robot Framework & \textit{Duplicated Keyword} \\
    Robot Framework & \textit{Duplicated Variable} \\
    Robot Framework & \textit{Hardcoded Environment} \\
    Robot Framework & \textit{Hiding Test Data} \\
    Robot Framework & \textit{Implementation Dependent} \\
    Robot Framework & \textit{Inconsistent Hierarchy} \\
    Robot Framework & \textit{Inconsistent Wording} \\
    Robot Framework & \textit{Lack Of Early Feedback} \\
    Robot Framework & \textit{Lack Of Encapsulation} \\
    Robot Framework & \textit{Lifeless} \\
    Robot Framework & \textit{Long Test Steps} \\
    Robot Framework & \textit{Middle Man} \\
    Robot Framework & \textit{Missing Assertion} \\
    Robot Framework & \textit{Missing Definition} \\
    Robot Framework & \textit{Narcissistic} \\
    Robot Framework & \textit{Noisy Logging} \\
    Robot Framework & \textit{On The Fly} \\
    Robot Framework & \textit{Over-Checking} \\
    Robot Framework & \textit{Pointless Scenario Descriptions} \\
    Robot Framework & \textit{Sensitive Locators} \\
    Robot Framework & \textit{Sneaky Checking} \\
    Robot Framework & \textit{Stinky Synchronization} \\
    Robot Framework & \textit{Test Data Loss} \\
    Robot Framework & \textit{Testing Data Not Code} \\
    Robot Framework & \textit{Transitive Dependency} \\
    Robot Framework & \textit{Unnecessary Navigation} \\
    Robot Framework & \textit{Unrealistic Data} \\
    Robot Framework & \textit{Unsecured Test Data} \\
    Robot Framework & \textit{Unsuitable Naming} \\
    Smaltalk & \textit{Magic Literals} \\
    TTCN-3 & \textit{Activation Asymmetry} \\
    TTCN-3 & \textit{Bad Comment Rate} \\
    TTCN-3 & \textit{Bad Documentation Comment} \\
    TTCN-3 & \textit{Complex Conditional} \\
    TTCN-3 & \textit{Constant Actual Parameter Value} \\
    TTCN-3 & \textit{Disorder} \\
    TTCN-3 & \textit{Duplicate Alt Branches} \\
    TTCN-3 & \textit{Duplicate Component Definition} \\
    TTCN-3 & \textit{Duplicate In-Line Templates} \\
    TTCN-3 & \textit{Duplicate Local Variable/Constant/Timer} \\
    TTCN-3 & \textit{Duplicate Statements} \\
    TTCN-3 & \textit{Duplicate Template Fields} \\
    TTCN-3 & \textit{Duplicated Code In Conditional} \\
    TTCN-3 & \textit{Fully-Parametrised Template} \\
    TTCN-3 & \textit{Goto} \\
    TTCN-3 & \textit{Idle Ptc} \\
    TTCN-3 & \textit{Insufficient Grouping} \\
    TTCN-3 & \textit{Long Parameter List} \\
    TTCN-3 & \textit{Long Statement Block} \\
    TTCN-3 & \textit{Magic Values} \\
    TTCN-3 & \textit{Missing Log} \\
    TTCN-3 & \textit{Missing Variable Definition} \\
    TTCN-3 & \textit{Missing Verdict} \\
    TTCN-3 & \textit{Nested Conditional} \\
    TTCN-3 & \textit{Over-Specificrunson} \\
    TTCN-3 & \textit{Short Template} \\
    TTCN-3 & \textit{Singular Component Variable/Constant/Timer Reference} \\
    TTCN-3 & \textit{Singular Template Reference} \\
    TTCN-3 & \textit{Stop In Function} \\
    TTCN-3 & \textit{Unreachable Default} \\
    TTCN-3 & \textit{Unrestricted Imports} \\
    TTCN-3 & \textit{Unused Definition} \\
    TTCN-3 & \textit{Unused Imports} \\
    TTCN-3 & \textit{Unused Parameter} \\
    TTCN-3 & \textit{Unused Variable Definition} \\
    TTCN-3 & \textit{Wasted Variable Definition} \\
    Typescript/Jest & \textit{Empty Tests} \\
    Typescript/Jest & \textit{Timeouts} \\
\end{longtable}
\end{scriptsize}

\section{Test Smells Refactorings by Language and Framework}
\label{ap:test-smells-refactorings}

\begin{scriptsize}
\begin{longtable}{p{3cm}p{6cm}p{3cm}p{1cm}}
    \caption{Test Smells Refactorings by Language and Framework} \\
    \toprule
    \textbf{Test Smell} & \textbf{Refactoring} & \textbf{Language/Framework} & \textbf{Papers} \\
    \midrule
    \endfirsthead
    \toprule
    \textbf{Test Smell} & \textbf{Refactoring} & \textbf{Language/Framework} & \textbf{Papers} \\
    \midrule
    \endhead
    \bottomrule
    \endfoot
    \midrule
    \endlastfoot
    \textit{Anonymous Test} & Rename method & Java/JUnit & S55 \\
    \textit{Army Of Clones} & Ensure that the user keyword that is called by a test and have at least one clone in the test suite is removed & Robot Framework & S83 \\
    \textit{Arrange\&quit} & Replace the if-return block by an assume for the precondition & Java/JUnit & S114 \\
    \textit{Assert Pre-condition} & Replace the assert pre-condition by the junit assume & Java/JUnit & S114 \\
    \textit{Assertion In Loop} & Create test cases for the edge cases and vulnerable values & Java & S1 \\
    \textit{Assertion In Loop} & Utilize repeated test feature & Java & S1 \\
    \textit{Assertion Roulette} & Add assertion explanation & Java & S2 \\
    \textit{Assertion Roulette} & Add assertion explanation & Java/JUnit & S23 \\
    \textit{Assertion Roulette} & Add assertion explanation & Java/JUnit & S26 \\
    \textit{Assertion Roulette} & Add assertion explanation & Java/JUnit & S27 \\
    \textit{Assertion Roulette} & Add assertion explanation & Java/JUnit & S28 \\
    \textit{Assertion Roulette} & Add assertion explanation & Java/JUnit & S37 \\
    \textit{Assertion Roulette} & Add assertion explanation & Scala/ScalaTest & S48 \\
    \textit{Assertion Roulette} & Add assertion explanation & Java/JUnit & S49 \\
    \textit{Assertion Roulette} & Add assertion explanation & Java/JUnit & S5 \\
    \textit{Assertion Roulette} & Add assertion explanation & Java/JUnit & S60 \\
    \textit{Assertion Roulette} & Add assertion explanation & Java/JUnit & S7 \\
    \textit{Assertion Roulette} & Add assertion explanation & Java & S73 \\
    \textit{Assertion Roulette} & Add assertion explanation & Java/JUnit & S75 \\
    \textit{Assertion Roulette} & Add assertion explanation & Java/JUnit & S86 \\
    \textit{Assertion Roulette} & Add assertion explanation & Java/JUnit & S93 \\
    \textit{Assertion Roulette} & Add assertion explanation & Java/JUnit & S104 \\
    \textit{Assertion Roulette} & Extract method & Java/JUnit & S23 \\
    \textit{Assertion Roulette} & Extract method & Java/JUnit & S55 \\
    \textit{Assertion Roulette} & Extract method & Java/JUnit & S7 \\
    \textit{Assertion Roulette} & Extract method & Java/JUnit & S75 \\
    \textit{Assertion Roulette} & Extract method & Python/Unittest & S84 \\
    \textit{Assertion Roulette} & Extract method & Python/Nose & S84 \\
    \textit{Assertion Roulette} & Extract method & Python/PyTest & S84 \\
    \textit{Assertion Roulette} & Single-condition tests & Java/JUnit & S26 \\
    \textit{Assertion Roulette} & Single-condition tests & Java/JUnit & S5 \\
    \textit{Assertion Roulette} & Surround assertions with assertall & Java/JUnit & S75 \\
    \textit{Assertion Roulette} & Use junit5 assertall & Java/JUnit & S107 \\
    \textit{Assertion Roulette} & Use junit5 grouped assertions & Java/JUnit & S26 \\
    \textit{Assertion Roulette} & Wrap assert on withclue & Scala/ScalaTest & S48 \\
    \textit{Bad Naming} & Rename test classes & Java/JUnit & S75 \\
    \textit{Conditional Assertions} & Ensure that the conditional assertion node is removed from the call graph & Robot Framework & S83 \\
    \textit{Conditional Test Logic} & Add guarded assertion & Java/JUnit & S26 \\
    \textit{Conditional Test Logic} & Encapsulating this test logic in a test utility method with an intent-revealing name & Java/JUnit & S26 \\
    \textit{Conditional Test Logic} & Extract method & Java/JUnit & S5 \\
    \textit{Conditional Test Logic} & Extract method & Java/JUnit & S7 \\
    \textit{Conditional Test Logic} & Refactoring tests into separate modules (e.g., one for each operating system), making each test module fully executable on the target os & Python/PyTest & S35 \\
    \textit{Conditional Test Logic} & Refactoring tests into separate modules (e.g., one for each operating system), making each test module fully executable on the target os & Python/Unittest & S35 \\
    \textit{Conditional Test Logic} & Transforming os-specific tests that check the os in the test code into their test decorators counterparts, whenever possible & Python/PyTest & S35 \\
    \textit{Conditional Test Logic} & Transforming os-specific tests that check the os in the test code into their test decorators counterparts, whenever possible & Python/Unittest & S35 \\
    \textit{Conditional Test Logic} & Use junit 5 conditional test execution & Java/JUnit & S26 \\
    \textit{Conditional Test Logic} & Use junit 5 repeated tests feature & Java/JUnit & S26 \\
    \textit{Dead Field} & Extract class & Java/JUnit & S54 \\
    \textit{Dead Field} & Extract class & Java/TestNG & S54 \\
    \textit{Duplicate Assert} & Extract method & Java & S2 \\
    \textit{Duplicate Assert} & Extract method & Java/JUnit & S26 \\
    \textit{Duplicate Assert} & Extract method & Java/JUnit & S27 \\
    \textit{Duplicate Assert} & Extract method & Java/JUnit & S28 \\
    \textit{Duplicate Assert} & Extract method & Java/JUnit & S37 \\
    \textit{Duplicate Assert} & Extract method & Java/JUnit & S7 \\
    \textit{Duplicate Assert} & Extract method & Java/JUnit & S104 \\
    \textit{Duplicate Assert} & Use parameterized test & Java/JUnit & S23 \\
    \textit{Duplicate Assert} & Use parameterized test & Java/JUnit & S26 \\
    \textit{Duplicate Assert} & Use parameterized test & Java/JUnit & S104 \\
    \textit{Duplicate Test Runs Caused By Inheritance} & Avoid inheritance from non-abstract test cases & Java/JUnit & S56 \\
    \textit{Duplicate Test Runs Caused By Inheritance} & Avoid inheritance from non-abstract test cases & Java/TestNG & S56 \\
    \textit{Duplicated Test Code} & Extract method & Java/JUnit & S17 \\
    \textit{Duplicated Test Code} & Extract method & Java/JUnit & S26 \\
    \textit{Duplicated Test Code} & Extract method & Java/JUnit & S49 \\
    \textit{Duplicated Test Code} & Extract method & Java/JUnit & S5 \\
    \textit{Duplicated Test Code} & Extract method & Java/JUnit & S86 \\
    \textit{Duplicated Test Code} & Implicity setup & Java/JUnit & S18 \\
    \textit{Duplicated Test Code} & Use junit 5 repeated tests & Java/JUnit & S26 \\
    \textit{Duplicated Test Code} & Use parameterized test & Java/JUnit & S26 \\
    \textit{Eager Test} & Code addition & Java/JUnit & S75 \\
    \textit{Eager Test} & Code removal & Java/JUnit & S75 \\
    \textit{Eager Test} & Extract method & Java & S2 \\
    \textit{Eager Test} & Extract method & Java/JUnit & S25 \\
    \textit{Eager Test} & Extract method & Java/JUnit & S28 \\
    \textit{Eager Test} & Extract method & Java & S45 \\
    \textit{Eager Test} & Extract method & Scala/ScalaTest & S48 \\
    \textit{Eager Test} & Extract method & Java/JUnit & S49 \\
    \textit{Eager Test} & Extract method & Java/JUnit & S55 \\
    \textit{Eager Test} & Extract method & Java/JUnit & S60 \\
    \textit{Eager Test} & Extract method & Java & S73 \\
    \textit{Eager Test} & Extract method & Java/JUnit & S86 \\
    \textit{Eager Test} & Extract method & Java/JUnit & S104 \\
    \textit{Exception Handling} & Extract method & Java/JUnit & S7 \\
    \textit{Exception Handling} & Replace try/catch block with @test expected annotation & Java/JUnit & S75 \\
    \textit{Exception Handling} & Replace try/catch block, @test expected, or @rule annotations with the assertthrows & Java/JUnit & S75 \\
    \textit{Exception Handling} & Use ""teardown"" instead of ""finnaly"" & Java/JUnit & S5 \\
    \textit{Exception Handling} & Use @rule someexception e = someexception.none() & Java/JUnit & S5 \\
    \textit{Exception Handling} & Use assertthrows and assertexcpetion & Java/JUnit & S7 \\
    \textit{Exception Handling} & Use assumptions & Java/JUnit & S7 \\
    \textit{Exception Handling} & Use catchthrowable & Java/JUnit & S5 \\
    \textit{Exception Handling} & Use expected=myexception.class & Java/JUnit & S5 \\
    \textit{Exception Handling} & Use unit 5 exception handling feature & Java/JUnit & S26 \\
    \textit{Exception Handling} & Using framework-specific features to handle exception testing & Java/JUnit & S26 \\
    \textit{For Testers Only} & Extract subclass & Java/JUnit & S86 \\
    \textit{For Testers Only} & Move method & Java/JUnit & S86 \\
    \textit{General Fixture} & Creating a minimal fixture, which covers only the setup code common for all test methods & Java/JUnit & S54 \\
    \textit{General Fixture} & Creating a minimal fixture, which covers only the setup code common for all test methods & Java/TestNG & S54 \\
    \textit{General Fixture} & Extract class & Java & S45 \\
    \textit{General Fixture} & Extract method & Java & S45 \\
    \textit{General Fixture} & Extract method & Java & S73 \\
    \textit{General Fixture} & Extract method & Java/JUnit & S86 \\
    \textit{General Fixture} & Extract method & Scala/ScalaTest & S48 \\
    \textit{General Fixture} & Inline method & Java/JUnit & S86 \\
    \textit{General Fixture} & Remove fixture & Scala/ScalaTest & S48 \\
    \textit{General Fixture} & Removing trait beforeandafter from the test class and demoting the fields referenced in the argument to method before to local, immutable variables in the appropriate test case & Scala/ScalaTest & S48 \\
    \textit{General Fixture} & Splitting the fixture into multiple smaller fixtures & Scala/ScalaTest & S13 \\
    \textit{General Fixture} & Splitting the fixture into multiple smaller fixtures & Scala/ScalaTest & S48 \\
    \textit{Hardcoded Environment} & Ensure the hardcoded argument in a call to a library keyword annotated as "configuration" is replaced with a variable & Robot Framework & S83 \\
    \textit{Hiding Test Data} & Ensure the call to the library keywords annotated as "getter" in the setup of test t is removed & Robot Framework & S83 \\
    \textit{Ignored Test} & Code removal & Java/JUnit & S75 \\
    \textit{Inappropriate Assertion} & Replace reserved words with an appropriate assertion & Java/JUnit & S75 \\
    \textit{Inappropriate Assertion} & Replace the not operator within the assertions & Java/JUnit & S75 \\
    \textit{Inappropriate Assertion} & Split conditional expressions into two different parameters & Java/JUnit & S75 \\
    \textit{Indirect Testing} & Extract method & Java/JUnit & S49 \\
    \textit{Indirect Testing} & Extract method & Java/JUnit & S60 \\
    \textit{Indirect Testing} & Extract method & Java/JUnit & S86 \\
    \textit{Indirect Testing} & Move method & Java/JUnit & S49 \\
    \textit{Indirect Testing} & Move method & Java/JUnit & S60 \\
    \textit{Interacting Test Suites} & Use junit 5 temporary directory & Java/JUnit & S26 \\
    \textit{Lack Of Cohesion Of Test Methods} & Extract class & Java & S45 \\
    \textit{Lack Of Cohesion Of Test Methods} & Extract class & Java/JUnit & S54 \\
    \textit{Lack Of Cohesion Of Test Methods} & Extract class & Java/TestNG & S54 \\
    \textit{Lack Of Cohesion Of Test Methods} & Extract method & Java & S45 \\
    \textit{Lack Of Encapsulation} & Ensure the direct call to a library keyword is removed from the acceptance criteria & Robot Framework & S83 \\
    \textit{Lazy Test} & Inline method & Java/JUnit & S86 \\
    \textit{Lazy Test} & Merge the individual test cases that execute the same method into a single one & Scala/ScalaTest & S48 \\
    \textit{Lazy Test} & Use parameterized test & Java/JUnit & S26 \\
    \textit{Long Test Steps} & Ensure the number of actions performed on the sut by a long step sees its value pass under the threshold l & Robot Framework & S83 \\
    \textit{Magic Number} & Replace magic literal & Java & S2 \\
    \textit{Magic Number} & Replace magic literal & Java/JUnit & S5 \\
    \textit{Magic Number} & Replace magic literal & Java & S65 \\
    \textit{Magic Number} & Replace magic literal & Java/JUnit & S104 \\
    \textit{Magic Number} & Replace magic literal & Java/JUnit & S107 \\
    \textit{Middle Man} & Ensure the delegate keyword call is replaced with another call which is not simply delegating its actions & Robot Framework & S83 \\
    \textit{Missing Assertion} & Ensure library keyword annotated as "assertion" is introduced in test that is missing any assertion & Robot Framework & S83 \\
    \textit{Multiple Aaa} & Extract method & Java/JUnit & S114 \\
    \textit{Multiple Acts} & Break the test case into separate ones, each focusing on one act and add separate assert & Java/JUnit & S114 \\
    \textit{Mystery Guest} & Inline resource & Java/JUnit & S86 \\
    \textit{Mystery Guest} & Manage resources explicitly in a fixture & Scala/ScalaTest & S48 \\
    \textit{Mystery Guest} & Setup external resource & Java/JUnit & S26 \\
    \textit{Mystery Guest} & Setup external resource & Java/JUnit & S49 \\
    \textit{Mystery Guest} & Setup external resource & Java/JUnit & S60 \\
    \textit{Mystery Guest} & Setup external resource & Java & S73 \\
    \textit{Mystery Guest} & Setup external resource & Java/JUnit & S86 \\
    \textit{Mystery Guest} & Use junit 5 temporary directory & Java/JUnit & S26 \\
    \textit{Narcissistic} & Ensure the name of a symptomatic "test steps" is changed so that it does not contain a personal pronoun anymore & Robot Framework & S83 \\
    \textit{Non-deterministic Data Test} & Adopt hashing algorithm to generate deterministic randomness & Java & S1 \\
    \textit{Non-deterministic Data Test} & Create a class and store n random data beforehand and then hand them out without synchronization & Java & S1 \\
    \textit{Non-deterministic Data Test} & Set a seed number for random generators to produce deterministic random data & Java & S1 \\
    \textit{Non-functional Statement} & Remove the empty scope or implement the expected functionality & Python/PyTest & S8 \\
    \textit{Non-functional Statement} & Remove the empty scope or implement the expected functionality & Python/Unittest & S8 \\
    \textit{Obscure Assert} & Eliminate unnecessary control flow to simply the logic of assert & Java/JUnit & S114 \\
    \textit{Obscure In-line Setup} & Extract method & Java/JUnit & S68 \\
    \textit{Obscure In-line Setup} & Extract method & Java/TestNG & S68 \\
    \textit{Obscure In-line Setup} & If the in-line setup is common to all tests, one can use an implicit setup & Java/JUnit & S54 \\
    \textit{Obscure In-line Setup} & If the in-line setup is common to all tests, one can use an implicit setup & Java/TestNG & S54 \\
    \textit{Obscure In-line Setup} & Setup code be moved into delegate setup methods & Java/JUnit & S54 \\
    \textit{Obscure In-line Setup} & Setup code be moved into delegate setup methods & Java/TestNG & S54 \\
    \textit{Parameter Order Wrong} & Correct the wrong parameter order in assertions & Java & S1 \\
    \textit{Poor Test Organization} & Separate source and test code and use parallel package structures & Java & S1 \\
    \textit{Programming Paradigms Blend} & Fields located outside the test class can be initialized in the setup method or in a helper method inside the test class & Python/PyTest & S8 \\
    \textit{Programming Paradigms Blend} & Fields located outside the test class can be initialized in the setup method or in a helper method inside the test class & Python/Unittest & S8 \\
    \textit{Redundant Print} & Code removal & Java/JUnit & S75 \\
    \textit{Resource Leakage} & Use junit 5 temporary directory & Java/JUnit & S26 \\
    \textit{Resource Optimism} & Add mock & Java/JUnit & S7 \\
    \textit{Resource Optimism} & Check before use & Java/JUnit & S7 \\
    \textit{Resource Optimism} & Create the resource in the setup & Java/JUnit & S7 \\
    \textit{Resource Optimism} & Make resource unique & Java/JUnit & S86 \\
    \textit{Resource Optimism} & Setup external resource & Java/JUnit & S49 \\
    \textit{Resource Optimism} & Setup external resource & Java/JUnit & S5 \\
    \textit{Resource Optimism} & Setup external resource & Java/JUnit & S60 \\
    \textit{Resource Optimism} & Setup external resource & Java/JUnit & S86 \\
    \textit{Resource Optimism} & Use junit 5 temporary directory & Java/JUnit & S26 \\
    \textit{Resource Optimism} & Use temp files (in case the resource is a file) & Java/JUnit & S7 \\
    \textit{Return In Test} & Remove the return statement in test cases & Java & S1 \\
    \textit{Scattered Test Fixtures Caused By Inheritance} & Manage test fixtures individually and independently in each test case & Java/JUnit & S56 \\
    \textit{Scattered Test Fixtures Caused By Inheritance} & Manage test fixtures individually and independently in each test case & Java/TestNG & S56 \\
    \textit{Scattered Test Fixtures Caused By Inheritance} & Use junit's @rule annotation to refactor the code that needs to be executed before and after a test & Java/JUnit & S56 \\
    \textit{Scattered Test Fixtures Caused By Inheritance} & Use junit's @rule annotation to refactor the code that needs to be executed before and after a test & Java/TestNG & S56 \\
    \textit{Scattered Test Fixtures Caused By Inheritance} & Use test utility classes to manage test fixtures instead of using inheritance & Java/JUnit & S56 \\
    \textit{Scattered Test Fixtures Caused By Inheritance} & Use test utility classes to manage test fixtures instead of using inheritance & Java/TestNG & S56 \\
    \textit{Sensitive Equality} & Compare the members of the object states structurally instead of relying on tostring() of the wholes & Scala/ScalaTest & S48 \\
    \textit{Sensitive Equality} & Introduce equality method & Java & S2 \\
    \textit{Sensitive Equality} & Introduce equality method & Java/JUnit & S49 \\
    \textit{Sensitive Equality} & Introduce equality method & Java/JUnit & S5 \\
    \textit{Sensitive Equality} & Introduce equality method & Java/JUnit & S60 \\
    \textit{Sensitive Equality} & Introduce equality method & Java & S73 \\
    \textit{Sensitive Equality} & Introduce equality method & Java/JUnit & S86 \\
    \textit{Sleepy Test} & Add mock & Java/JUnit & S7 \\
    \textit{Sleepy Test} & Code addition & Java/JUnit & S75 \\
    \textit{Sleepy Test} & Refactor the test smell using java's future library & Java & S65 \\
    \textit{Sleepy Test} & Reorder tests & Java/JUnit & S7 \\
    \textit{Sleepy Test} & Use intelligent waiting library & Java/JUnit & S7 \\
    \textit{Sleepy Test} & Use waitfor() condition in the java awaitility library & Java & S65 \\
    \textit{Suppressed Exception} & Remove the catch and keep the try & Java/JUnit & S114 \\
    \textit{Test Maverick} & Extract class & Java/JUnit & S54 \\
    \textit{Test Maverick} & Extract class & Java/TestNG & S54 \\
    \textit{Test Run War} & Make resource unique & Java/JUnit & S26 \\
    \textit{Test Run War} & Make resource unique & Java/JUnit & S86 \\
    \textit{Test Run War} & Reorder tests & Java/JUnit & S7 \\
    \textit{Test Run War} & Use junit 5 resource lock feature & Java/JUnit & S26 \\
    \textit{Undefined Test} & Rename the test method using the prefix "test" or remove it & Python/PyTest & S8 \\
    \textit{Undefined Test} & Rename the test method using the prefix "test" or remove it & Python/Unittest & S8 \\
    \textit{Unknown Test} & Add assertion & Java/JUnit & S7 \\
    \textit{Unknown Test} & Add assertion & Java/JUnit & S114 \\
    \textit{Unknown Test} & Code addition & Java/JUnit & S75 \\
    \textit{Unknown Test} & Remove method & Java/JUnit & S7 \\
    \textit{Using Test Case Inheritance To Test Source Code Polymorphism} & Use parameterized test & Java/JUnit & S56 \\
    \textit{Using Test Case Inheritance To Test Source Code Polymorphism} & Use parameterized test & Java/TestNG & S56 \\
    \textit{Vague Header Setup Smell} & Field initializations should be placed in an implicit setup to specify the behavior and the places to inspect within a class & Java/JUnit & S54 \\
    \textit{Vague Header Setup Smell} & Field initializations should be placed in an implicit setup to specify the behavior and the places to inspect within a class & Java/TestNG & S54 \\
    \textit{Verifying In Setup Method} & Instead of using it in the setup method that runs before each test method, use it in the setupclass method that runs only once & Python/PyTest & S8 \\
    \textit{Verifying In Setup Method} & Instead of using it in the setup method that runs before each test method, use it in the setupclass method that runs only once & Python/Unittest & S8 \\
\end{longtable}
\end{scriptsize}

\section{Datasets referenced in RQ4\label{ap:rq4-datasets}}

\begin{table}[H]
\scriptsize
    \centering
    \caption{Test Smells Dataset Overview}
    \label{tab:test-smells-dataset}
    \begin{tabular}{p{2cm}p{1.5cm}p{9cm}}
        \toprule
        \textbf{Language/ Framework} & \textbf{N. of Test Smells} & \textbf{Dataset URL} \\
        \midrule
        C\# & 4 & \url{https://figshare.com/articles/dataset/Reinforcement_Learning_from_Static_Quality_Metrics/25983166?file=46868443} \\
        C\#/xUnit & 7 & \url{https://figshare.com/articles/dataset/A_Road_to_Find_Them_All_Towards_an_Agnostic_Strategy_for_Test_Smell_Detection/26444968} \\
        C\#/xUnit & 16 & \url{https://github.com/Partha-SUST16/xNose/tree/main/results} \\
        C++/GoogleTest & 7 & \url{https://figshare.com/articles/dataset/A_Road_to_Find_Them_All_Towards_an_Agnostic_Strategy_for_Test_Smell_Detection/26444968} \\
        Java & 16 & \url{https://github.com/bhpachulski/SAST21-Paper} \\
        Java & 4 & \url{https://github.com/darioamorosodaragona-tuni/ML-Test-Smell-Detection-Online-Appendix} \\
        Java/JUnit & 1 & \url{https://github.com/unittesting-nonpublic/private-keep-out_replication-package} \\
        Java/JUnit & 1 & \url{https://doi.org/10.5281/zenodo.11267987} \\
        Java/JUnit & 4 & \url{https://zenodo.org/records/3337892#.XswWby-w3yU} \\
        Java/JUnit & 7 & \url{https://figshare.com/articles/dataset/A_Road_to_Find_Them_All_Towards_an_Agnostic_Strategy_for_Test_Smell_Detection/26444968} \\
        Java/JUnit & 7 & \url{https://figshare.com/s/b1d6b70e10837aaf3f17?file=41800866} \\
        Java/JUnit & 19 & \url{https://testsmells.org/pages/testsmellexamples.html} \\
        Java/JUnit & 20 & \url{https://figshare.com/s/7b8bf9a7580001929f63?file=27729300} \\
        Java/JUnit & 21 & \url{https://figshare.com/s/da2b1903e7b209b1f77e?file=50726838} \\
        Python/Unittest & 14 & \url{https://figshare.com/s/4789bd212185042cdad0?file=49671705} \\
        Python/Unittest & 17 & \url{https://se.cite.ehime-u.ac.jp/data/Fushihara_SEAA2023/} \\
        Python/Unittest & 18 & \url{https://se.cite.ehime-u.ac.jp/data/Fushihara_SEAA2024/} \\
        Robot Framework & 16 & \url{https://github.com/kabinja/suit-smells-replication-package} \\
        \bottomrule
    \end{tabular}
\end{table}

\section{Detection and refactoring tools identified in RQ6\label{ap:rq6-tools}}

\begin{scriptsize}
\begin{longtable}[c]{p{1.5cm} p{2cm} p{3.5cm} p{2cm} p{1.5cm} p{1cm}}
    \caption{Detection and Refactoring Tools}%
    \label{tab:detecting-refactoring-tools}\\
    \toprule
    \textbf{Language} & \textbf{Framework} & \textbf{Tool} & \textbf{Capabilities} & \textbf{Availability} & \textbf{Papers} \\
    \midrule
    \endfirsthead
    \multicolumn{6}{c}{{\bfseries Table \thetable\ continued from previous page}} \\
    \toprule
    \textbf{Language} & \textbf{Framework} & \textbf{Tool} & \textbf{Capabilities} & \textbf{Availability} & \textbf{Papers} \\
    \midrule
    \endhead
    \midrule
    \multicolumn{6}{r}{{Continued on next page}} \\
    \endfoot
    \bottomrule
    \endlastfoot
    C\# & xUnit & XNose & Detection & \href{https://github.com/Partha-SUST16/xNose}{Available} & S99 \\
    Java & JUnit & Better Code Hub (Extension) & Detection & Unavailable & S34 \\
    Java & JUnit & DesigniteJava (Extension) & Detection & \href{https://www.designite-tools.com/products-dj}{Available} & S85 \\
    Java & JUnit & Evosuite extension & Detection & Unavailable & S40 \\
    Java & JUnit & JNose & Detection & \href{https://github.com/arieslab/jnose}{Available} & S11 \\
    Java & JUnit & JTDog & Detection & \href{https://github.com/kusumotolab/JTDog}{Available} & S33 \\
    Java & JUnit & RAIDE & Detection and Refactoring & \href{https://raideplugin.github.io/RAIDE/}{Available} & S23, S37 \\
    Java & JUnit & SimilarTestAnalysis & Detection & \href{https://bitbucket.org/felixfangzh/similartestanalysis/src/master/}{Available} & S17 \\
    Java & JUnit & SniffTest & Detection & \href{https://github.com/MaierFlorian/SniffTest}{Available} & S77 \\
    Java & JUnit & TASTE & Detection & Unavailable & S50 \\
    Java & JUnit & TestComet \& Teslo & Detection & Unavailable & S93 \\
    Java & JUnit & TESTLER & Detection and Refactoring & \href{https://github.com/saltlab/Testler}{Available} & S19 \\
    Java & JUnit & TSDETECT & Detection & \href{https://testsmells.org/}{Available} & S79 \\
    Java & JUnit & TSGame & Detection & Unavailable & S97 \\
    Java & JUnit, TestNG & FSE\_Ignore\_Test & Detection & \href{https://github.com/boyang9602/FSE\_Ignore\_Test}{Available} & S43 \\
    Java & JUnit, TestNG & TestHound & Detection & \href{http://swerl.tudelft.nl/bin/view/MichaelaGreiler/TestHound}{Available} & S54 \\
    Java & Selenium WebDriver & TEDD & Detection & \href{https://github.com/matteobiagiola/FSE19-submission-material-TEDD}{Available} & S21 \\
    Java & ~ & DARTS & Detection and Refactoring & \href{https://github.com/StefanoLambiase/DARTS}{Available} & S45 \\
    Java & ~ & DFD & Detection & \href{https://tinyurl.com/ybk29dss}{Available} & S72 \\
    Java & ~ & DFI & Detection & Unavailable & S80 \\
    Java & ~ & JNose & Detection & \href{https://github.com/arieslab/jnose}{Available} & S70 \\
    Java & ~ & Unamed & Detection & \href{https://github.com/Yanming-Yang/TestSmellPackage}{Available} & S1 \\
    JavaScript, Typescript & Chai, Jest & STEEL & Detection & Unavailable & S4 \\
    Python & Nose, PyTest, Unittest & Pysta & Detection & Unavailable & S84 \\
    Python & PyTest & PyTest-Smell & Detection & \href{https://pypi.org/project/pytest-smell/}{Available} & S14 \\
    Python & PyTest, Unittest & TEMPY & Detection & \href{https://github.com/danieldavidf/TEMPY}{Available} & S3 \\
    Python & Unittest & PYNOSE & Detection & \href{https://github.com/JetBrains-Research/PyNose}{Available} & S31 \\
    Scala & ScalaTest & SoCRATES & Detection & \href{https://github.com/jonas-db/socrates}{Available} & S13, S48 \\
    TTCN-3 & ~ & TRex & Detection and Refactoring & \href{https://www.trex.informatik.uni-goettingen.de/trac}{Available} & S89, S91 \\
    ~ & Robot Framework & sonar-ikora-plugin & Detection & \href{https://github.com/serval-uni-lu/sonar-ikora-plugin}{Available} & S83 \\
    ~ & Robot Framework & Ukwikora & Detection & \href{https://github.com/kabinja/ukwikora-inspector}{Available} & S78 \\
    C++, C\#, Java & GoogleText, xUnit, JUnit & SniffML & Detection & \href{https://github.com/daviromao/sniffer}{Available} & S102 \\
    Java & JUnit & UTRefactor's & Detection and Refactoring & \href{https://github.com/testmigrator/testsmellrefactoring}{Available} & S104 \\
    Java & JUnit & Sojourner under Sabotage & Detection & \href{https://sojourner-under-sabotage.se2.fim.uni-passau.de/}{Available} & S106 \\
    Java & JUnit & Viscount & Detection & \href{https://github.com/unittesting-nonpublic/viscount}{Available} & S111, S112 \\
    Python & PyTest & PyTeRor & Detection & \href{https://github.com/semaki2000/PyTeRor}{Available} & S117 \\
\end{longtable}
\end{scriptsize}

\section{Studies reporting test smell criticality (RQ5)\label{ap:rq5-criticality-studies}}

\begin{scriptsize}
\begin{longtable}[c]{p{3.6cm}p{1.4cm}p{7.0cm}}
    \caption{Studies reporting test smell criticality (RQ5)}
    \label{tab:rq5-criticality-by-study}\\
    \toprule
    \textbf{Test smell} & \textbf{Paper} & \textbf{Subject} \\
    \midrule
    \endfirsthead
    \multicolumn{3}{c}{{\bfseries Table \thetable\ continued from previous page}} \\
    \toprule
    \textbf{Test smell} & \textbf{Paper} & \textbf{Subject} \\
    \midrule
    \endhead
    \midrule
    \multicolumn{3}{r}{{Continued on next page}} \\
    \endfoot
    \bottomrule
    \endlastfoot
    \textit{Ambiguous Display Name} & S100 & Impact on usefulness in development tasks \\
    \textit{Assertion Roulette} & S2 & Developer perception in general \\
    \textit{Assertion Roulette} & S7 & Developer perception in general \\
    \textit{Assertion Roulette} & S24 & Maintainability \\
    \textit{Assertion Roulette} & S67 & Maintainability \\
    \textit{Assertion Roulette} & S16 & Not clear \\
    \textit{Assertion Roulette} & S81 & Not clear \\
    \textit{Conditional Test Logic} & S7 & Developer perception in general \\
    \textit{Conditional Test Logic} & S34 & Maintainability \\
    \textit{Conditional Test Logic} & S67 & Maintainability \\
    \textit{Constructor Initialization} & S34 & Maintainability \\
    \textit{Duplicate Assert} & S2 & Developer perception in general \\
    \textit{Duplicate Assert} & S7 & Developer perception in general \\
    \textit{Duplicate Assert} & S98 & Developer perception in general \\
    \textit{Duplicate Display Name} & S100 & Impact on usefulness in development tasks \\
    \textit{Eager Test} & S2 & Developer perception in general \\
    \textit{Eager Test} & S98 & Developer perception in general \\
    \textit{Eager Test} & S24 & Maintainability \\
    \textit{Eager Test} & S34 & Maintainability \\
    \textit{Eager Test} & S60 & Maintainability \\
    \textit{Eager Test} & S67 & Maintainability \\
    \textit{Eager Test} & S81 & Not clear \\
    \textit{Empty Test} & S98 & Developer perception in general \\
    \textit{Empty Test} & S67 & Maintainability \\
    \textit{Exception Handling} & S7 & Developer perception in general \\
    \textit{General Fixture} & S67 & Maintainability \\
    \textit{General Fixture} & S16 & Not clear \\
    \textit{Ignored Test} & S24 & Maintainability \\
    \textit{Ignored Test} & S67 & Maintainability \\
    \textit{Inconsistent Display Name} & S100 & Impact on usefulness in development tasks \\
    \textit{Inconsistent Style} & S100 & Impact on usefulness in development tasks \\
    \textit{Lazy Test} & S98 & Developer perception in general \\
    \textit{Magic Number} & S2 & Developer perception in general \\
    \textit{Magic Number} & S98 & Developer perception in general \\
    \textit{Magic Number} & S24 & Maintainability \\
    \textit{Magic Number} & S67 & Maintainability \\
    \textit{Magic Number} & S16 & Not clear \\
    \textit{Missing Display Name} & S100 & Impact on usefulness in development tasks \\
    \textit{Multilingual Display Name} & S100 & Impact on usefulness in development tasks \\
    \textit{Mystery Guest} & S24 & Maintainability \\
    \textit{Mystery Guest} & S34 & Maintainability \\
    \textit{Mystery Guest} & S67 & Maintainability \\
    \textit{Non-functional Statement} & S8 & Not clear \\
    \textit{Obscure Test} & S34 & Maintainability \\
    \textit{Obscure Test} & S67 & Maintainability \\
    \textit{Programming Paradigms Blend} & S8 & Not clear \\
    \textit{Redundant Assertion} & S34 & Maintainability \\
    \textit{Redundant Display Name} & S100 & Impact on usefulness in development tasks \\
    \textit{Resource Optimism} & S7 & Developer perception in general \\
    \textit{Resource Optimism} & S24 & Maintainability \\
    \textit{Resource Optimism} & S34 & Maintainability \\
    \textit{Resource Optimism} & S67 & Maintainability \\
    \textit{Sensitive Equality} & S34 & Maintainability \\
    \textit{Sleepy Test} & S7 & Developer perception in general \\
    \textit{Sleepy Test} & S24 & Maintainability \\
    \textit{Sleepy Test} & S34 & Maintainability \\
    \textit{Sleepy Test} & S67 & Maintainability \\
    \textit{Sleepy Test} & S16 & Not clear \\
    \textit{Test Run War} & S7 & Developer perception in general \\
    \textit{Unclear Display Name} & S100 & Impact on usefulness in development tasks \\
    \textit{Undefined Test} & S8 & Not clear \\
    \textit{Ungrammatical Display Name} & S100 & Impact on usefulness in development tasks \\
    \textit{Unknown Test} & S7 & Developer perception in general \\
    \textit{Unknown Test} & S16 & Not clear \\
    \textit{Verifying In Setup Method} & S8 & Not clear \\
\end{longtable}
\end{scriptsize}

\section{Test smells outside the Open Catalog: manual category assignments\label{ap:outside-catalog-categories}}

\begin{scriptsize}
\begin{longtable}[c]{p{5.8cm}p{5.8cm}}
    \caption{Test smells outside the Open Catalog with manually assigned top-level category.}
    \label{tab:outside-catalog-categories}\\
    \toprule
    \textbf{Test smell} & \textbf{Assigned category} \\
    \midrule
    \endfirsthead
    \multicolumn{2}{c}{{\bfseries Table \thetable\ continued from previous page}} \\
    \toprule
    \textbf{Test smell (normalized)} & \textbf{Assigned category} \\
    \midrule
    \endhead
    \midrule
    \multicolumn{2}{r}{{Continued on next page}} \\
    \endfoot
    \bottomrule
    \endlastfoot
    \textit{Complicated Setup Scenarios} & Issues in test steps \\
    \textit{Dependencies On Test Containers} & Dependencies \\
    \textit{Duplicated Setup} & Issues in test steps \\
    \textit{Friction} & Code related \\
    \textit{Implementation Dependent} & Test semantic - logic \\
    \textit{Lack Of Early Feedback} & Issues in test steps \\
    \textit{Private Method Test} & Test semantic - logic \\
    \textit{State Leak} & Dependencies \\
    \textit{Test Data Loss} & Issues in test steps \\
    \textit{Testing Data Not Code} & Test semantic - logic \\
    \textit{Transitive Dependency} & Dependencies \\
    \textit{Unrealistic Data} & Issues in test steps \\
    \textit{Timeouts} & Test execution - behavior \\
    \textit{Arrange\&Quit} & Issues in test steps \\
    \textit{Testing Only Field Accessors} & Test semantic - logic \\
    \textit{Time Acquisition} & Test execution - behavior \\
    \textit{Unsecured Test Data} & Issues in test steps \\
    \textit{Ambiguous Display Name} & Code related \\
    \textit{Annotation Repetition} & Code related \\
    \textit{Code Organization} & Code related \\
    \textit{Coupling With Production Code} & Code related \\
    \textit{Directly Executing Ui Scripts} & Code related \\
    \textit{Duplicate Check} & Code related \\
    \textit{Duplicate Display Names} & Code related \\
    \textit{Duplicate Test Runs Caused By Inheritance} & Code related \\
    \textit{Duplicated Keyword} & Code related \\
    \textit{Duplicated Variable} & Code related \\
    \textit{God Test Class} & Code related \\
    \textit{Goto} & Code related \\
    \textit{Hard To Co-Evolve} & Code related \\
    \textit{Inconsistent Display Name} & Code related \\
    \textit{Inconsistent Style} & Code related \\
    \textit{Magic Literals} & Code related \\
    \textit{Magic Number} & Code related \\
    \textit{Missing Definition} & Code related \\
    \textit{Missing Display Name} & Code related \\
    \textit{Mock Call Verification} & Code related \\
    \textit{Mock Makes The Test Always Passes} & Code related \\
    \textit{Mock Naming} & Code related \\
    \textit{Multilingual Display Name} & Code related \\
    \textit{Naming Convention Violation} & Code related \\
    \textit{Obsolete Tests} & Code related \\
    \textit{Poor Test Organization} & Code related \\
    \textit{Redudant Test} & Code related \\
    \textit{Redundant Display Name} & Code related \\
    \textit{Redundant Test} & Code related \\
    \textit{Replicating Disadvantages Of The Code} & Code related \\
    \textit{Test Clone} & Code related \\
    \textit{Tooling Details In Test Case} & Code related \\
    \textit{Unclear Display Name} & Code related \\
    \textit{Unecessary Test} & Code related \\
    \textit{Ungrammatical Display Name} & Code related \\
    \textit{Using More Than One Mock For A Test} & Code related \\
    \textit{Dependencies Between Tests} & Dependencies \\
    \textit{Dependent Assert} & Dependencies \\
    \textit{Exceptions Due To External Dependencies} & Dependencies \\
    \textit{Lack Of Cohesion} & Dependencies \\
    \textit{Resource Optimist} & Dependencies \\
    \textit{Parameter Order Wrong} & Design related \\
    \textit{Assert Pre-Condition} & Issues in test steps \\
    \textit{Asserting Constants} & Issues in test steps \\
    \textit{Asserting Object Initialization Multiple Times} & Issues in test steps \\
    \textit{Assertion In Loop} & Issues in test steps \\
    \textit{Assertions With Not Related Parent Class Method} & Issues in test steps \\
    \textit{Exceptions Due To Incomplete Setup} & Issues in test steps \\
    \textit{Exceptions Due To Null Arguments} & Issues in test steps \\
    \textit{Inappropriate Assertion} & Issues in test steps \\
    \textit{Missing Assert} & Issues in test steps \\
    \textit{Missing Assertion} & Issues in test steps \\
    \textit{Not Asserted Return Values} & Issues in test steps \\
    \textit{Not Asserted Side Effects} & Issues in test steps \\
    \textit{Obscure Assert} & Issues in test steps \\
    \textit{Redundant Not Null Assertion} & Issues in test steps \\
    \textit{Scattered Test Fixtures Caused By Inheritance} & Issues in test steps \\
    \textit{Suboptimal Assert} & Issues in test steps \\
    \textit{Suppressed Exception} & Issues in test steps \\
    \textit{Unrelated Assertions} & Issues in test steps \\
    \textit{Vague Header Setup Smell} & Issues in test steps \\
    \textit{Multithreading} & Test execution - behavior \\
    \textit{Happy Path Only} & Test semantic - logic \\
    \textit{Return In Test} & Design related \\
    \textit{Testing The Same Exception Scenario} & Code related \\
    \textit{Testing The Same Void Method} & Code related \\
    \textit{Hardcoded Dependency} & Code related \\
    \textit{Pointless Scenario Descriptions} & Design related \\
    \textit{Multiple Aaa} & Design related \\
    \textit{Multiple Acts} & Design related \\
    \textit{Data Creep} & Issues in test steps \\
    \textit{Ordered Hypothesis} & Test semantic - logic \\
    \textit{Lifeless} & Code related \\
    \textit{Doppelg\"anger} & Code related \\
\end{longtable}
\end{scriptsize}

\FloatBarrier






\end{document}